\documentclass{tlp}
\usepackage{epsfig}
\usepackage[utf8]{inputenc}
\usepackage{multicol}
\usepackage{multirow}
\usepackage{wrapfig}
\usepackage{subfigure}
\usepackage{graphicx}
\graphicspath{./figures/}

\begin{document}

\label{firstpage}

\title[Threads and Or-Parallelism Unified]
      {Threads and Or-Parallelism Unified}

\author[Vítor Santos Costa, Inês Dutra and Ricardo Rocha]
       {VíTOR SANTOS COSTA, INÊS DUTRA and RICARDO ROCHA\\
       CRACS \& INESC-Porto LA, Faculty of Sciences, University of Porto\\
       Rua do Campo Alegre, 1021/1055, 4169-007 Porto, Portugal\\
       \email{\{vsc,ines,ricroc\}@dcc.fc.up.pt}}

\maketitle

\begin{abstract}
  One of the main advantages of Logic Programming (LP) is that it
  provides an excellent framework for the parallel execution of
  programs. In this work we investigate novel techniques to
  efficiently exploit parallelism from real-world applications in low
  cost multi-core architectures. To achieve these goals, we revive and
  redesign the YapOr system to exploit or-parallelism based on a
  multi-threaded implementation. Our new approach takes full advantage
  of the state-of-the-art fast and optimized YAP Prolog engine and
  shares the underlying execution environment, scheduler and most of
  the data structures used to support YapOr's model. Initial
  experiments with our new approach consistently achieve almost linear
  speedups for most of the applications, proving itself as a good
  alternative for exploiting implicit parallelism in the currently
  available low cost multi-core architectures.
\end{abstract}



\section{Introduction}

One of the main advantages of Logic Programming (LP) is that it
provides an excellent framework for the parallel execution of
programs. \emph{Implicit} parallelism occurs naturally in logic
programs; or-parallelism arises because different alternatives can be
run independently; and and-parallelism arises when different goals can
be run in different processors~\cite{Gupta-01}. On the other hand, as
a very high-level language, LP has often been used (and is still
indeed being used) to \emph{explicitly} control parallelism and manage
tasks~\cite{Fonseca-09}. It is therefore unsurprising that parallelism
has been an important subject in the development of LP.

Arguably, the high-point of parallel LP were the early
nineties. During this period, a number of very sophisticated and
well-engineered systems were built and shown to be quite successful at
exploiting implicit or-parallelism~\cite{Aurora-88,Ali-90a,Gupta-99},
implicit and-parallelism~\cite{Hermenegildo-91,Shen-92,Pontelli-97},
and even a combination of both~\cite{CostaVS-91}. The hard lesson was
that although these systems did deliver performance, they never became
widely used. We believe this can be explained for a number of reasons:
\textbf{(i)} only few users had access to the expensive parallel
machines of the day; \textbf{(ii)} they required significant changes
to the Prolog engine, thus becoming complex to maintain and install;
and \textbf{(iii)} large Prolog programs are hard to
parallelize. Thus, interest in the whole area of declarative parallel
programming laid dormant for a number of years.

The increasing availability and popularity of multi-core processors
has changed this equation: multi-core systems have become a viable
high-performance, affordable and standardized alternative to the
traditional (and often expensive) parallel architectures. In fact, as
the number of cores per processor continues to increase, interest in
parallelism expands. This has led to a recent revival of research in
the area and motivates a simple question: \emph{Is implicit
  parallelism worth it in multi-core architectures?}

Answering this question requires a parallel LP system. Unfortunately,
building parallel logic programming systems from scratch is
\emph{hard}. Ideally, we would like to reuse the vast amount of
preexisting work. But, will systems designed and built more than a
decade ago still work on the much faster modern architectures? And can
we integrate them in modern Prolog (or LP) engines?

We believe that the answer to this question lies in recent progress on
supporting \emph{multi-threading} in Prolog engines. Motivated by the
desire to support concurrent applications and explicit parallelism in
systems such as Ciao~\cite{Carro-99} and
SWI-Prolog~\cite{Wielemaker-03}, multi-threading has now become widely
available in Prolog engines, such as YAP~\cite{CostaVS-08} and XSB
Prolog~\cite{Marques-10}. Nowadays, supporting threads can be seen as
a requirement of modern Prolog engines, not an extension.

To demonstrate the feasibility of implicit parallelism in thread-based
systems, we revive and redesign the YapOr system~\cite{Rocha-99b} to
exploit or-parallelism based on a multi-threaded implementation. YapOr
is an or-parallel engine, originally based on the stack copying model,
as first implemented in the Muse system~\cite{Ali-90a}, that extends
the YAP Prolog system to support implicit or-parallelism in logic
programs.  Our new approach takes full advantage of YAP's
state-of-the-art fast and optimized engine and shares the underlying
execution environment, scheduler and most of the data structures used
to support parallelism in YapOr.  Our new design thus unifies YAP's
multi-threaded support with YapOr's or-parallelism support. We named
our new approach the \emph{Threads and Or-parallelism unified (ThOr)}
model.

To put the performance of our new implementation in perspective, we
experimented with the system on a number of different computing
platforms and compare it against the copy based YapOr. Initial
experiments with both systems consistently achieve almost linear
speedups for a large number of applications, and good speedups even if
parallelism is very fine-grained. On the other hand, ThOr benefits
from its simpler architecture and we show ThOr running on two
platforms where porting YapOr would be non-trivial. Thus, we believe
it is as an excellent alternative for exploiting implicit parallelism
in a portable way in the currently available low cost multi-core
architectures.

The remainder of the paper is organized as follows. First, we briefly
introduce the YapOr model and describe its main data structures,
memory organization and scheduler strategies. Then, we present our new
multi-threaded implementation, describe its major implementation
decisions and discuss its advantages, disadvantages and challenges. At
last, we present experimental results and we end by outlining some
conclusions.


\section{The YapOr Model}

The initial implementation of or-parallelism in YapOr was largely
based on the \emph{stack copying model} as first introduced by Ali and
Karlson in the Muse system~\cite{Ali-90a,Ali-90b}. YapOr is an example
of a \emph{multi-sequential} model~\cite{Ali-86}. In this approach,
each processor or \emph{worker} maintains its own copy of the search
tree where it is expected to spend most of its time performing
reductions. Only when a worker runs out of work, it searches for work
from fellow workers. If a fellow worker has work, it can make some or
all of its open alternatives available: this operation is called
\emph{sharing}. First, the sharer will make some or all of its
choice-points \emph{public}, so that backtracking to these
choice-points can be synchronized between different workers. Second,
in a copying model, the execution stacks of the sharer are copied to
the requester. The sharer then continues forward execution, while the
requester backtracks to the shared choice-points and exploits
alternatives.

Deciding which workers to ask for work and how much work should be
received is a function of the \emph{scheduler}. For stack copying,
scheduling strategies based on bottommost dispatching of work have
proved to be more efficient than topmost
strategies~\cite{Ali-90b}. Synchronization between workers is mainly
done through choice-points. In an copying model, each worker has a
separate copy of the public choice-points. Synchronization requires an
auxiliary data structure, called \emph{or-frame}, to be associated
with the public choice-points. We next discuss in more detail some of
these features and characteristics of the original YapOr model.

\subsection{Or-Frames and Public Choice-Points}

\begin{figure}
\centering
\includegraphics[width=11.5cm]{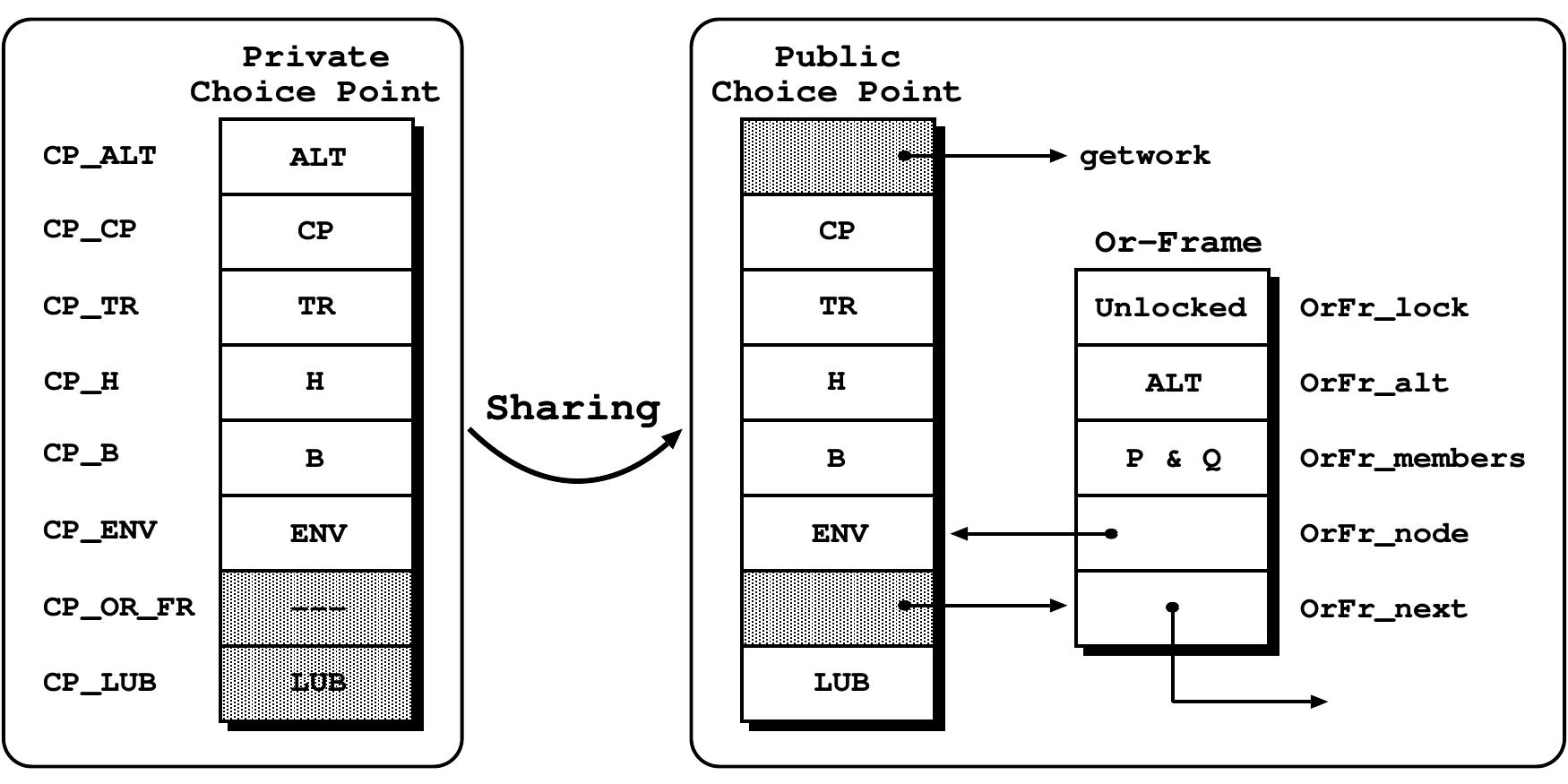}
\caption{Sharing a choice-point}
\label{fig_sharing_choicepoint}
\end{figure}

In order to correctly exploit a shared branch, a fundamental task when
sharing work is to turn public the private choice-points. Public
choice-points are treated differently because we need to synchronize
workers in such a way that we avoid executing twice the same
alternative. To do so, the worker sharing work adds an \emph{or-frame}
data structure to each private choice-point made public. The or-frames
form a tree that represents the public search
tree. Figure~\ref{fig_sharing_choicepoint} illustrates the relation
between the choice-points before and after that operation.

In YapOr, a choice-point includes eight fields, the first six were
inherited from YAP and the last two were introduced by YapOr. The
\texttt{CP\_OR\_FR} field points to the corresponding or-frame, if the
choice-point is public. Otherwise, it is not used. The
\texttt{CP\_LUB} field (LUB stands for ``Local Untried Branches'')
stores the number of private unexplored alternatives in the current
branch, and it is used to compute the worker's load. Sharing a
choice-point involves updating the \texttt{CP\_OR\_FR} and
\texttt{CP\_ALT} field, respectively, to the newly created or-frame
and to the \texttt{getwork} pseudo-instruction. Backtracking to a
shared choice-point will thus always trigger the execution of the
\texttt{getwork} instruction and its execution allows for a
synchronized access to the untried alternatives among the workers
sharing the corresponding or-frame. The or-frame is initialized as
follows. The \texttt{OrFr\_lock} field supports a busy-wait locking
mechanism that guarantees atomic updates to the or-frame data. It is
initially set to unlocked. The \texttt{OrFr\_alt} field stores the
\texttt{ALT} pointer which was previously in the {\tt CP\_ALT}
choice-point field (i.e., the control of the untried alternatives
moves to the or-frame). The workers sharing the choice-point are
marked in the \texttt{OrFr\_members} field. \texttt{OrFr\_node} is a
back pointer to the corresponding choice-point. Last, the
\texttt{OrFr\_next} field is a pointer to the parent or-frame on the
current branch.

\subsection{Incremental Copying}
\label{sec:inccopy}

Sharing work is a major source of overhead in YapOr, as it requires
copying the execution stacks between workers. The \emph{incremental
  copying strategy}~\cite{Ali-90a} is designed to reduce this overhead
by allowing the receiving worker to keep those parts of its execution
stacks that are consistent with the giving worker, and only to copy
the differences between the two workers' stacks.

For example, consider that worker \emph{Q} asks worker \emph{P} for
sharing and that worker \emph{P} decides to share its private nodes
with \emph{Q}. To implement incremental copying, \emph{Q} should start
by backtracking to the youngest common node with \emph{P}, therefore
becoming partially consistent with part of \emph{P}. Then, if \emph{Q}
receives a positive answer from \emph{P}, it only needs to copy the
differences between \emph{P} and \emph{Q}. These differences can be
easily calculated through the information stored in the common node
found by \emph{Q} and in the top registers of the local, heap and
trail stacks of \emph{P}. Care must be taken about variables older
than the common youngest node that were bound by worker $P$, as
incremental copying does not copy these bindings. Worker \emph{Q}
needs to explicitly \emph{install} the bindings for such
variables. This process, called the \textit{adjustment of cells
  outside the increments}, is implemented by searching the trail stack
for bindings to variables older than the common node~\cite{Ali-90a}.

\subsection{Memory Organization}

The YapOr memory is divided into two major \emph{shared} address
spaces: the \emph{global space} and a collection of \emph{local
  spaces}. The global space contains the code area inherited from YAP
and all the data structures necessary to support parallelism. Each
local space represents one system worker and contains the four
execution stacks inherited from YAP: heap, local, trail, and auxiliary
stack.

In order to efficiently meet the requirements of incremental copy, the
set of local spaces are mapped as follows. The starting worker asks
for shared memory in the system's initialization phase. Afterwards,
the remaining workers are created and inherit the previously mapped
address space. Then, each new worker rotates the local spaces, in such
a way that all workers will see their own spaces at the same virtual
memory addresses.

\begin{figure}
\centering
\includegraphics[width=11.7cm]{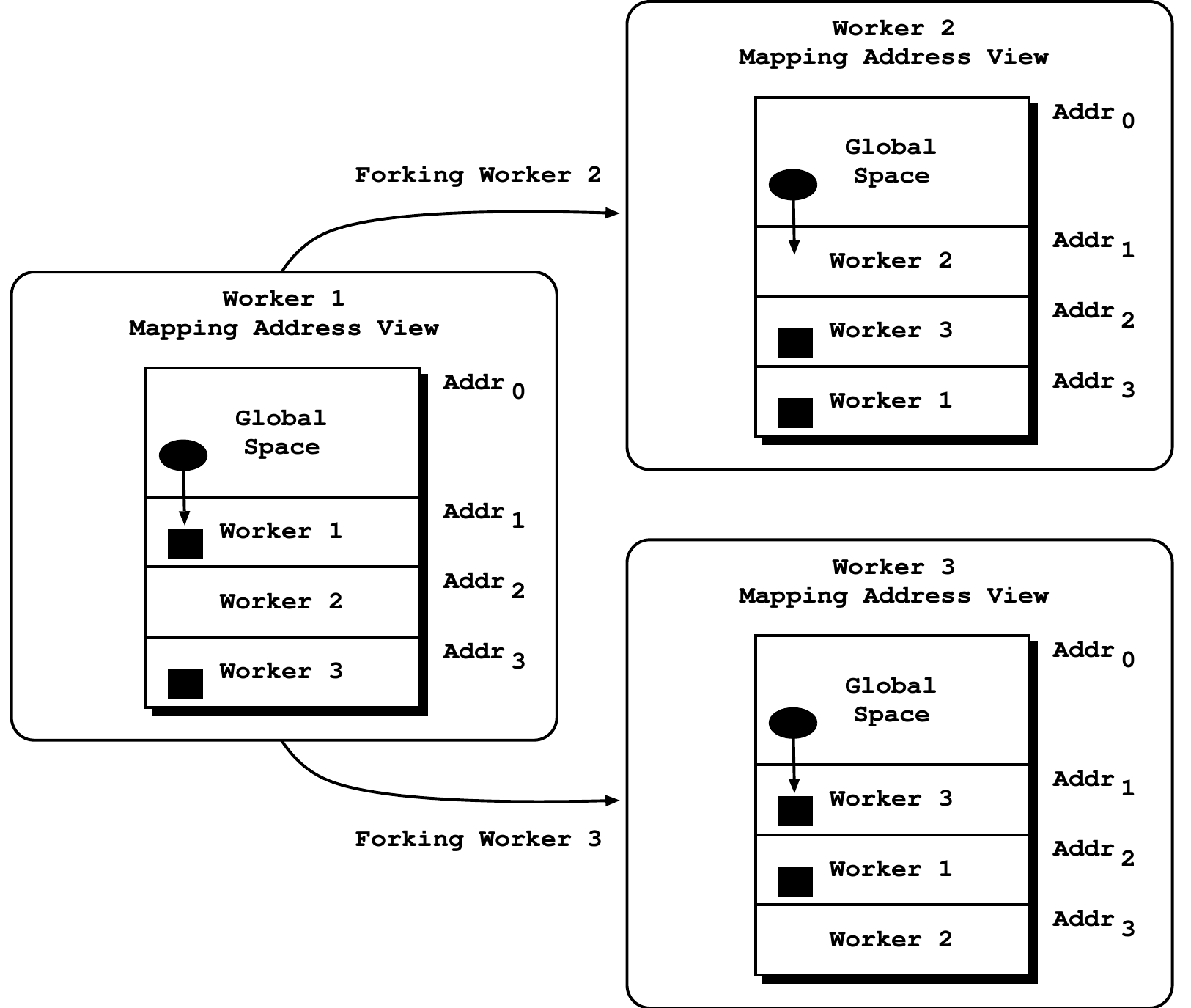}
\caption{YapOr's memory organization for 3 workers}
\label{fig_yapor_mem}
\end{figure}

Figure~\ref{fig_yapor_mem} helps to understand this remapping
scheme. It considers 3 workers and it illustrates the resulting
mapping address view of each worker after rotating the inherited local
spaces. The global space is shared at the same virtual address
($Addr_0$ in Fig.~\ref{fig_yapor_mem}) for the 3 workers and each
worker accesses its own local space starting from the same virtual
address ($Addr_1$ in Fig.~\ref{fig_yapor_mem}). The figure also shows
an example or-frame (as an oval) allocated in the global space and the
corresponding choice-points (as squares) for the workers sharing it,
workers 1 and 3. Notice that this virtual memory trickery allows the
same or-frame to point to different copies of the same choice-points,
depending on the worker. This mapping scheme also allows for efficient
memory copying operations during incremental copying because no
reallocation of address values in the copied segments is necessary. To
copy a stack segment between two workers, we simply copy directly from
one worker space to the relative virtual memory address in the other
worker's space.

In YapOr, this memory scheme is implemented through two different and
alternative UNIX shared memory management functionalities, the
\emph{mmap()} and \emph{shmget()} functions~\cite{Stevens-92}. These
functions allow us to map shared memory segments at given addresses,
and unmap and remap them later at new addresses.

\subsection{Scheduling Strategies}

When a worker runs out of work, first the scheduler tries to select a
busy worker with excess of work load to share work. There are two
alternatives to search for busy workers in the search tree: search
below or search above the current node. Idle workers always start to
search below the current node, and only if they do not find any busy
worker there, they search above. The main advantage of selecting a
busy worker below instead of above is that the idle worker can request
immediately the sharing operation, because its current node is already
common to the busy worker, which avoids backtracking in the tree and
undoing variable bindings.

When the scheduler does not find any busy worker with excess of work
load, it tries to move the idle worker to a better position in the
search tree. By default, the idle worker backtracks until it reaches a
node where there is at least one busy worker below. Another option is
to backtrack until reaching the node that contains all the busy
workers below. The goal of these strategies is to distribute the idle
workers in such a way that the probability of finding, as soon as
possible, busy workers with excess of work below is substantially
increased.

Although other memory organization models have performed well for
smaller numbers of workers, stack copying has consistently shown to
provide better scalability, while achieving low overheads.


\section{The ThOr Model}

The ThOr model builds upon two major components of YAP: the YapOr
implementation, as discussed in the previous section, and the threads
library~\cite{CostaVS-08}. The YAP thread library can be seen as a
high-level interface to the POSIX threads library, where each thread
runs on a separate stack but shares access to the global data
structures (code area, atom table and predicate table). As each thread
operates its own stack, thus, it is natural to \emph{associate} each
parallel worker to a thread: threads can run in parallel and they
already include the machinery to support shared access and updates to
the global data structures and input/output structures.

Notice however that not all threads have to be workers: some threads
may be used to support specialized tasks, possibly running in parallel
with the workers. We believe this is an important advantage of
ThOr. Traditionally, we expect to find a single \emph{or-parallel}
program, and issues such as side-effects must be addressed within the
or-parallel system. In ThOr, we can easily construct independent
threads that \emph{collaborate} or even \emph{control} the workers
(or-parallel threads). Natural applications are the collection of
solutions, and performing input/output tasks.

\subsection{Memory Model}

Three different or-parallel models have been implemented for YAP: the
Sparse Binding Array (SBA) model~\cite{Correia-97}, the $\alpha$-COW
model~\cite{CostaVS-99a} and YapOr's copying model. Of the three
models, the $\alpha$-COW relies on process forking and is therefore
not suitable for a thread-based implementation. This leaves us with
two choices: models sharing as much data structures as possible, such
as the SBA; and copying based models, such as the default YapOr
model. We chose to use copying for ThOr for several reasons:

\begin{enumerate}
\item Copying allows us to preserve a key notion in the thread
  library: independent and separate workers have a \emph{private}
  stack. In this way, we can reuse the existing code for threads so
  that workers can independently perform garbage collection and stack
  shifting.
\item Because workers see a contiguous stack, copying imposes less
  overheads on the engine and has high performance compared to the
  other approaches~\cite{CostaVS-00}.
\item Ultimately copying is less intrusive on the sequential
  engines. As a small experiment, we explored what kind of changes
  would be needed in the emulator. In both models, support for
  or-parallelism, including copying, requires about 60 changes to the
  emulator, mostly in order to adapt choice-point manipulation and to
  perform pruning on the shared tree. Support for the SBA model
  requires 90 additional changes that affect a complex operation,
  unification.
\end{enumerate}

On the other hand, YapOr's copying model relies on every worker
\emph{having its own stacks at the same virtual address
  position}. This clearly will not work with threads. Next, we discuss
how we have addressed this problem.

\subsection{Shifted Copying}

In a thread-based model, all memory areas should be visible to all
threads \emph{at the same virtual memory address positions}. Moreover,
in order to take full advantage of memory (especially on 32 bit
machines), it is convenient not to assume any preconditions on memory
organization. In YAP, threads may actually move their stacks in the
virtual memory space during execution.

The $\alpha$-COW, SBA and copying models share most of the scheduling
and work management code and all models assume that every worker has
its own stacks at the same virtual memory addresses. ThOr has also
been designed to take advantage of the existing code-base but, having
each own stack at the same virtual memory addresses, does not hold
true in ThOr. Namely, to share work we need to have several copies of
the same choice-point at different virtual memory addresses. To
address this problem, our key idea is \emph{shifted copying}, which
essentially consists of two steps: copy the memory between the workers
sharing work and then adjust pointers. Although shifted copying adds a
linear overhead to copying operations, it offers some important
advantages:

\begin{enumerate}
\item it allows using the thread infra-structure as is;
\item it allows shifting between stacks \emph{with different sizes},
  and we can actually reuse preexisting code from the YAP stack
  shifter.
\end{enumerate}

\begin{figure}
\centering
\includegraphics[width=5.5cm]{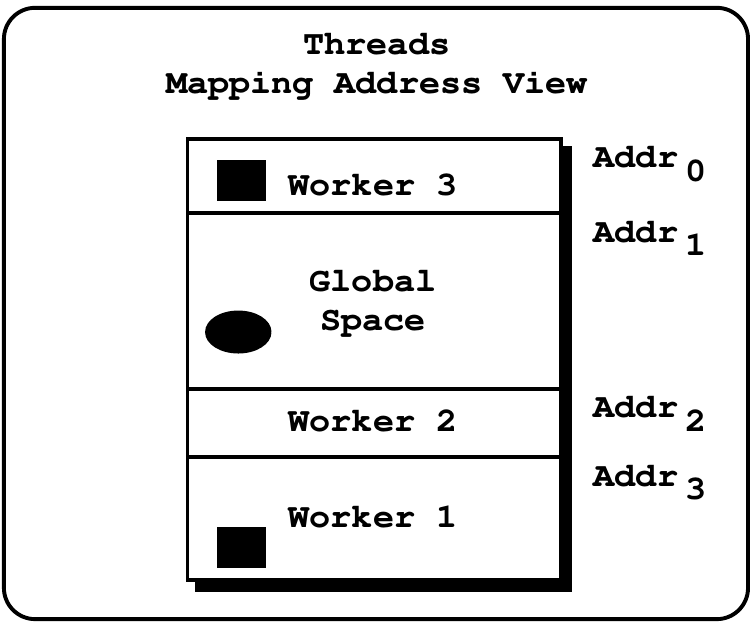}
\caption{ThOr's memory organization for 3 workers}
\label{fig_thor_mem}
\end{figure}

The resulting memory model for ThOr is presented in
Fig.~\ref{fig_thor_mem}. Notice that we do not assume any fixed order
addresses between stacks, and that we do not assume equal stack
sizes. Initialization code now consists of creating $N-1$ or-parallel
threads (and it is in fact simpler than the initialization code for
the other models). Choice-points in ThOr are stored as \emph{offsets}
from the top of stack: ThOr takes advantage of the fact that YAP does
not perform garbage collection on the local stack. Translation between
addresses and offsets is performed by a \emph{setter} and a
\emph{getter} method. As in YapOr, one can copy the whole stacks, or
perform \emph{incremental copying}. The latter is quite often much
more efficient, and we discuss it in some more detail next.

\subsection{Work Sharing Algorithm}

We next describe the work sharing algorithm used in ThOr. We assume
that worker $P$ gives work to worker $Q$. The algorithm for $P$ is
shown in Fig.~\ref{fig_giving_work} and the algorithm for $Q$ is shown
in Fig.~\ref{fig_receiving_work}. Notice that the \emph{signal[]}
array is used to synchronize between the two workers and that, by
default, all entries are in $ready$ mode.

\begin{figure}[ht]
\begin{minipage}[b]{0.4\linewidth}
\textbf{P\_SHARE} $(p,q)$ \{

~~~$share\_private\_nodes(q)$

~~~$signal[q] = nodes\_shared$

~~~\textbf{wait\_until} $(signal[q] == ready)$

\}

~

~

~

~

~
\caption{Giving Work}
\label{fig_giving_work}
\end{minipage}
\hspace{0.5cm}
\begin{minipage}[b]{0.4\linewidth}
\textbf{Q\_SHARE} $(p,q)$ \{

~~~\textbf{wait\_until} $(signal[q] \neq ready)$

~~~$copy\_registers(p,q)$

~~~\textbf{if} $(INCREMENTAL\_COPY)$

~~~~~~$copy\_stacks(p,q,deltas)$

~~~~~~$copy\_trailed\_entries(q,p)$

~~~~~~$signal[q] = ready$

~~~~~~$adjust\_stacks(deltas)$

~~~\textbf{else}

~~~~~~$copy\_stacks(p,q,full\_stacks)$

~~~~~~$signal[q] = ready$

~~~~~~$adjust\_stacks(full\_stacks)$

~~~\textbf{endif}

\}
\caption{Receiving Work}
\label{fig_receiving_work}
\end{minipage}
\end{figure}

Worker $P$ uses $share\_private\_nodes()$ to make all its
choice-points public. In practice this entails creating a new or-frame
per choice-point and waiting for worker $Q$. Our implementation moves
most of the work to worker $Q$: this worker is the one who actually
copies the stacks and adjusts the cells in the stacks. Copying may be
done incrementally or the full stacks may be copied. By default, ThOr
uses incremental copy, but full copy may be required after a garbage
collection. Last, for incremental copy ThOr copies \emph{and adjusts}
cells outside the increments (mentioned in Section~\ref{sec:inccopy}),
as required by the copying model.

\subsection{ThOr in Practice}

We have so far discussed the main principles of ThOr implementation.
Next, we discuss some important practical questions that we should
address to make ThOr usable in practice.

\paragraph{Sequential and Parallel Predicates}
\label{sec:sequ-parall-pred}

One major burden for or-parallel systems has been the need to fully
support Prolog semantics. We believe we should follow a different
approach. In ThOr, or-parallel execution is a component in a
multi-threaded system, and should focus on executing pure Prolog
programs.

Therefore, ThOr will not impose sequential execution of system
built-ins. Notice that synchronization of side-effects was already
supported in YapOr, and it is still possible, just it is not the
default execution mode.

\paragraph{Locking}
\label{sec:locking}

ThOr allows for two types of locking: low-level high-performance code,
and POSIX threads locks. The low-level locks are busy-waiting locks
written in assembly. We should observe that the locking and scheduling
protocols used in YapOr rely extensively on busy-waiting. As a result,
workers that are waiting for new tasks will still consume CPU
resources. This is a reasonable solution for dedicated hardware, but
it is not acceptable for personal workstations. In this case, the
scheduling code must be rewritten so that idle workers will not hog
the CPU.

\paragraph{Global Variables}
\label{sec:global-variables}

The \texttt{nb\_} and \texttt{b\_} primitives maintain global
variables~\cite{Demoen-98b}, by associating a name, represented as a
Prolog atom, to a term. In the backtrackable version, or \texttt{b\_},
the association is discarded on backtracking; in the non-backtrackable
version, the association can be seen as a global property. These
primitives are quite widely used, e.g., they are crucial in the
implementation of constraint libraries.

The SWI-Prolog (and YAP) thread libraries deem global variables to be
\emph{thread-private}. Each thread maintains separate lists of
different global variables, and threads cannot access other thread's
global variables. The current implementation of ThOr follows this
approach: we simply keep global variables on the original thread. We
believe this approach does not correspond well with or-parallelism, as
each thread now logically implements a snapshot of the \emph{same}
stack:

\begin{itemize}
\item\texttt{b\_} variables are in fact just a different way of
  accessing stack objects, and should be copied as any other variables
  by the stack copying mechanism.
\item \texttt{nb\_} variables should be seen as shared between all
  threads that work together in or-parallel. This suggests that these
  variables should be stored in a common area.
\end{itemize}

In order to provide support for \texttt{nb\_} variables, a separate
data area for global variables should exist and be shared between
threads.


\section{Experimental Results}

Our performance evaluation focuses on three aspects:

\begin{itemize}
\item Performance of well known Prolog on the ThOr implementation.
\item Performance of ThOr in comparison with that of YapOr and
  sequential YAP.
\item Scalability of ThOr on a multi-core machine of reasonable size.
\end{itemize}

The following programs were used as benchmarks:

\begin{itemize}
\item \texttt{cubes}: a known benchmark taken from Tick's
  book~\cite{Tick-87}. This program implements the solution of a magic
  cube of size 7.
\item \texttt{fp}: this is an implementation of a floor plan design,
  originally implemented by Kovács~\cite{Kovacs-92}, which represents
  a real-world application. The task is to partition a floor rectangle
  according to a room request list consisting of room sizes and
  constraints such as rooms that face north or have natural light.
\item \texttt{ham}: checks if a given graph forms a Hamiltonian cycle.
\item \texttt{magic}: a solution to the 3x3x3 magic cube.
\item \texttt{map} and \texttt{mapbigger}: a solution for the map
  coloring problem using 4 colors. We used two maps representing
  diverse size and graph density. The smaller version has 10 nodes and
  the bigger version has 17 nodes.
\item \texttt{puzzle}: one version of sudoku where the diagonals must
  add up to the same amount.
\item \texttt{puzzle4x4}: a solution for a 4x4 maze.
\item \texttt{queens}: a solution for the n-queens problem using
  forward checking. The size of the board used in testing was 13x13.
\end{itemize}

We performed our experiments on 2 different machines. The first is an
Intel(R) Core(TM)2, quad-core CPU Q9450 with 4 GBytes of RAM and
running Mandriva Linux in 32-bit mode. The second machine is a Dell
Poweredge R905, 4 six-core Opteron 8425HE (2.1Ghz) with 64 GBytes of
RAM and running Fedora 12 in 64-bit mode.

Table~\ref{tab:time1wkr} shows the base execution times (running on a
single core), in seconds, for all benchmarks in both machines. Each
benchmark was run at least 20 times. The results shown in the
Table~\ref{tab:time1wkr} are the averages of these runs.  In this
Table, we also show the base execution times of sequential YAP. The
numbers between parentheses correspond to the overhead imposed by
YapOr or Thor when compared with sequential YAP.

\begin{table}[hbtp]
\caption{Average execution times, in seconds, for single core of YapOr
  and ThOr on a Linux quad-core desktop machine and on a Linux 4
  six-core server machine, compared with sequential YAP}
\begin{tabular}{l|rrr|rrr} \hline
\multirow{2}{*}{\bf Benchmark} &
\multicolumn{3}{c|}{\bf Quad-Core} &
\multicolumn{3}{c}{\bf 4 Six-Core} \\ 
& \multicolumn{1}{c}{\bf YAP} & \multicolumn{1}{c}{\bf ThOr} & \multicolumn{1}{c|}{\bf YapOr}
& \multicolumn{1}{c}{\bf YAP} & \multicolumn{1}{c}{\bf ThOr} & \multicolumn{1}{c}{\bf YapOr}\\ 
\hline \hline
\texttt{cubes}     &  0.11 &  0.11 (1.00) &  0.11 (1.00) &  0.20 &  0.23 (1.15) &  0.20 (1.00) \\
\texttt{fp}        &  1.47 &  2.36 (1.61) &  1.71 (1.16) &  2.51 &  3.29 (1.31) &  2.67 (1.06) \\
\texttt{ham}       &  0.15 &  0.29 (1.93) &  0.19 (1.27) &  0.33 &  0.46 (1.36) &  0.34 (1.03) \\
\texttt{magic}     & 25.07 & 27.55 (1.10) & 27.80 (1.11) & 40.29 & 48.88 (1.21) & 41.16 (1.02) \\
\texttt{map}       & 12.20 & 20.25 (1.66) & 14.60 (1.20) & 24.06 & 30.45 (1.26) & 23.94 (0.99) \\
\texttt{mapbigger} & 33.01 & 55.26 (1.67) & 39.63 (1.20) & 64.46 & 81.09 (1.25) & 65.90 (1.02) \\
\texttt{puzzle}    &  0.08 &  0.13 (1.63) &  0.08 (1.00) &  0.15 &  0.20 (1.34) &  0.17 (1.13) \\
\texttt{puzzle4x4} &  6.02 &  7.18 (1.19) &  6.47 (1.07) &  9.17 & 10.34 (1.12) &  9.38 (1.02) \\
\texttt{queens}    & 21.79 & 24.54 (1.13) & 24.12 (1.11) & 48.10 & 51.63 (1.07) & 48.73 (1.01) \\
\hline
\emph{Average}     &       &       (1.44) &       (1.12) &       &       (1.23) &       (1.03) \\
\hline
\end{tabular}
\label{tab:time1wkr}
\end{table}

Table~\ref{tab:time1wkr} shows a higher significant distance between
base execution times of YapOr and ThOr on the quad-core desktop than
the distance observed on the 4 six-core server. One important
difference between the two machines is that the quad-core is running
in 32-bit mode, and the 4 six-core in 64-bit mode. In general, YAP has
been better optimized for 32-bit mode, and the overhead of supporting
YapOr is more noticeable in this mode. Namely, the parallel code
always tests for work requests, even if there is a single worker.

In the 4 six-core server, the YapOr version almost has no overheads in
these benchmarks. These overheads range from 0 (for \texttt{map},
where the performances of Yap and YapOr are practically the same plus
or minus a rounding error) to 13\% (for \texttt{puzzle}).  ThOr incurs
a higher overhead that ranges from 7 to 36\%. This is partly because
threads need to disable some optimizations in YAP (namely, abstract
machine arguments cannot be at fixed addresses). But, most
importantly, threads require indirect access to thread-private global
variables, including the abstract machine registers. This overhead can
be reduced by ensuring that every function has a local variable
pointing to the abstract machine, but the optimization is currently
only implemented in the emulator, but not in the built-in predicates
compiled inline.

\begin{figure}
\subfigure[\tt cubes]{\label{fig:cubes_4}
          \hspace{-1cm}\includegraphics[width=6cm]{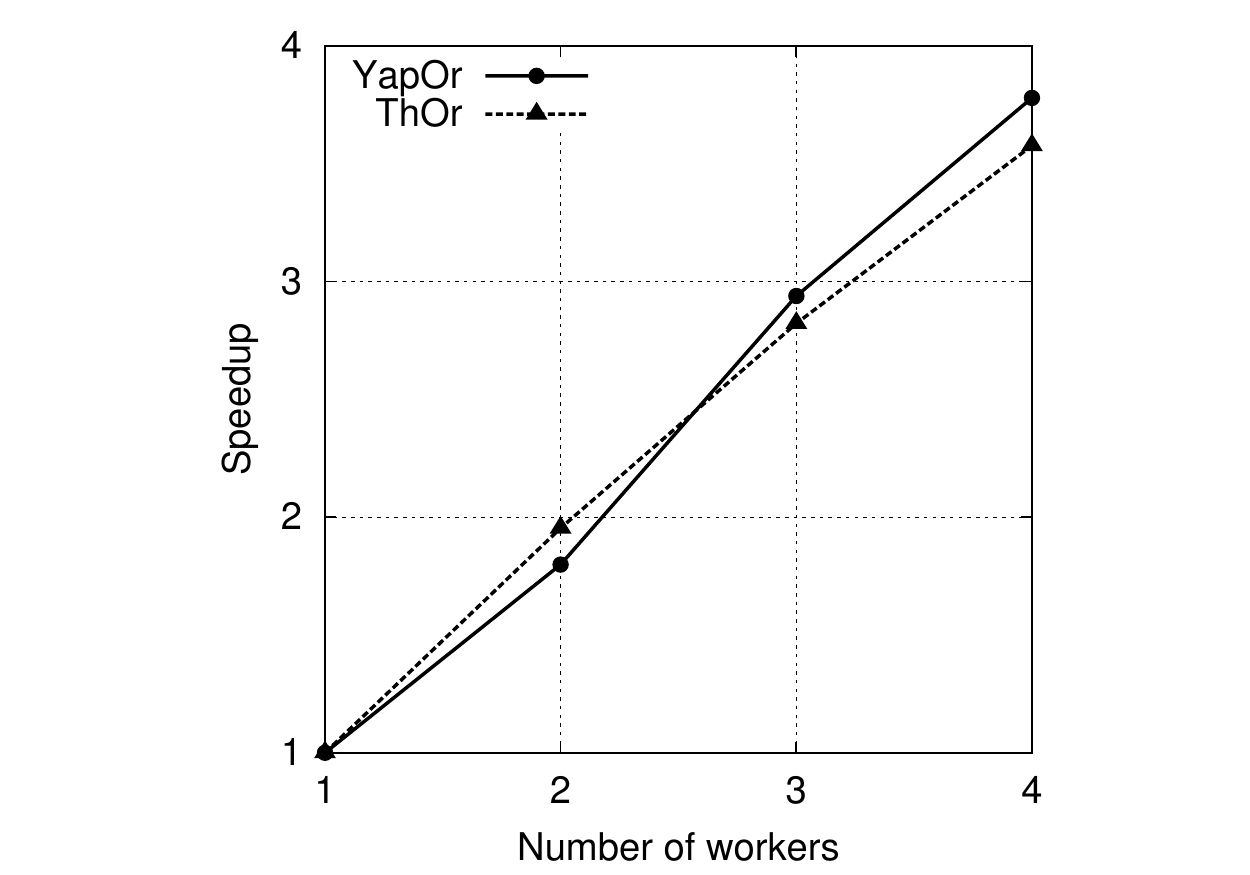}\hspace{-1cm}}
\subfigure[\tt fp]{\label{fig:fp_4}
          \hspace{-1cm}\includegraphics[width=6cm]{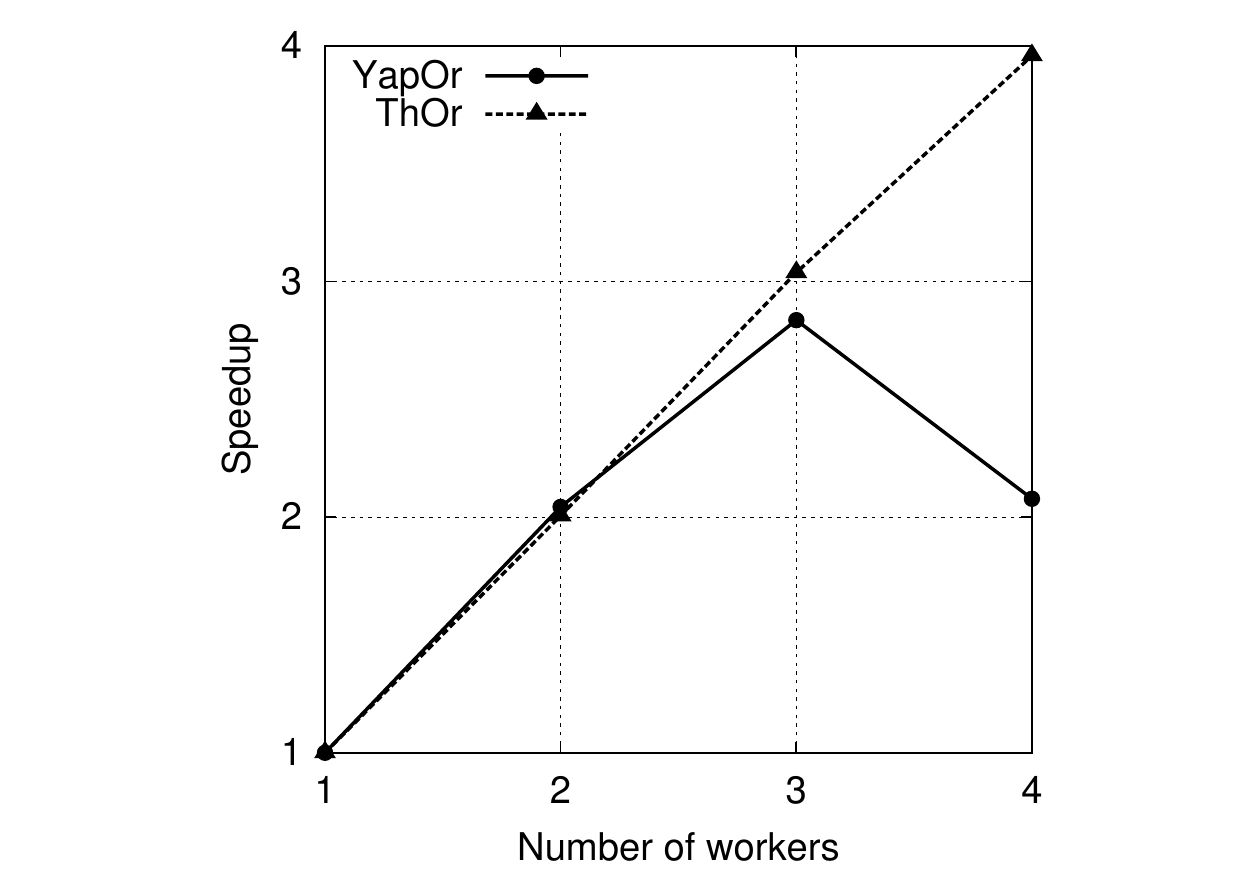}\hspace{-1cm}}
\subfigure[\tt ham]{\label{fig:ham_4}
          \hspace{-1cm}\includegraphics[width=6cm]{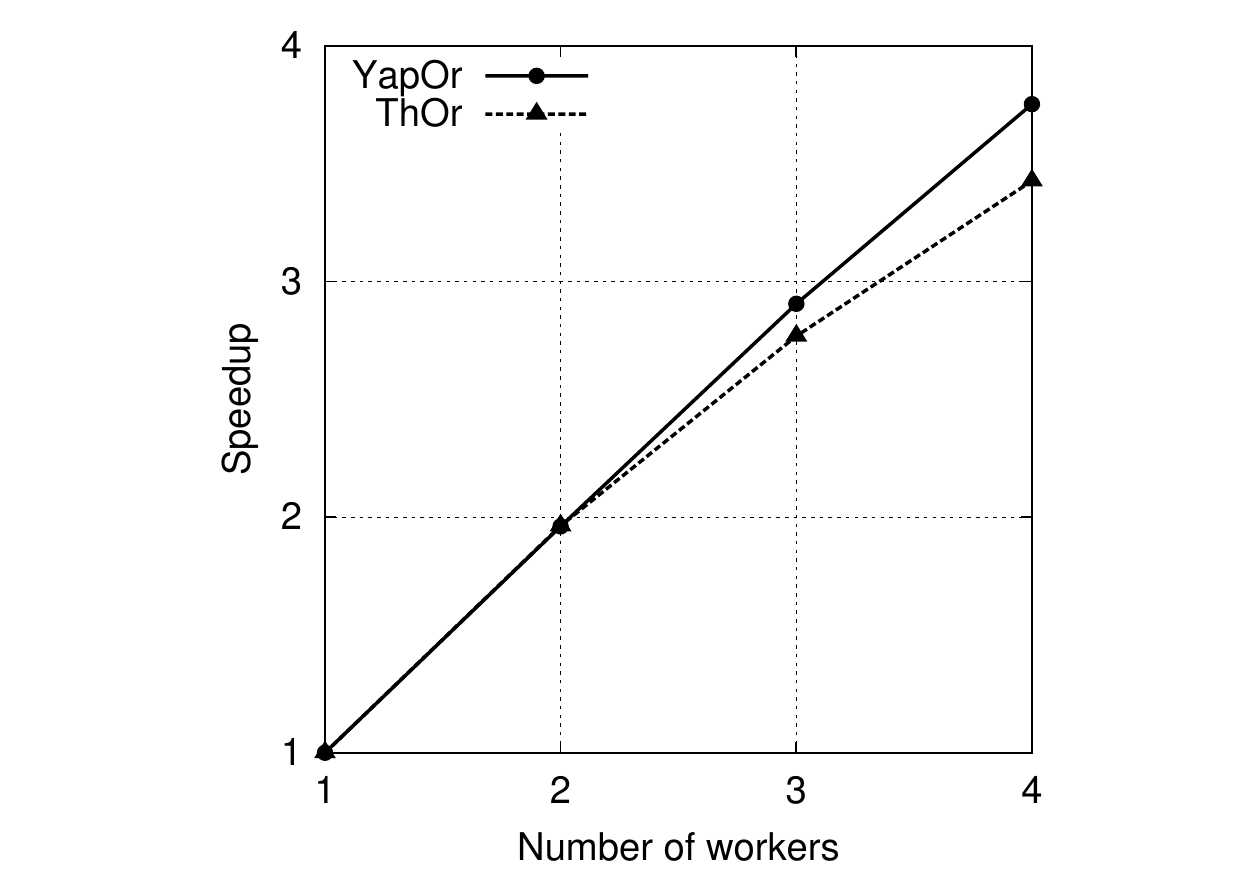}\hspace{-1cm}}
\subfigure[\tt magic]{\label{fig:magic_4}
          \hspace{-1cm}\includegraphics[width=6cm]{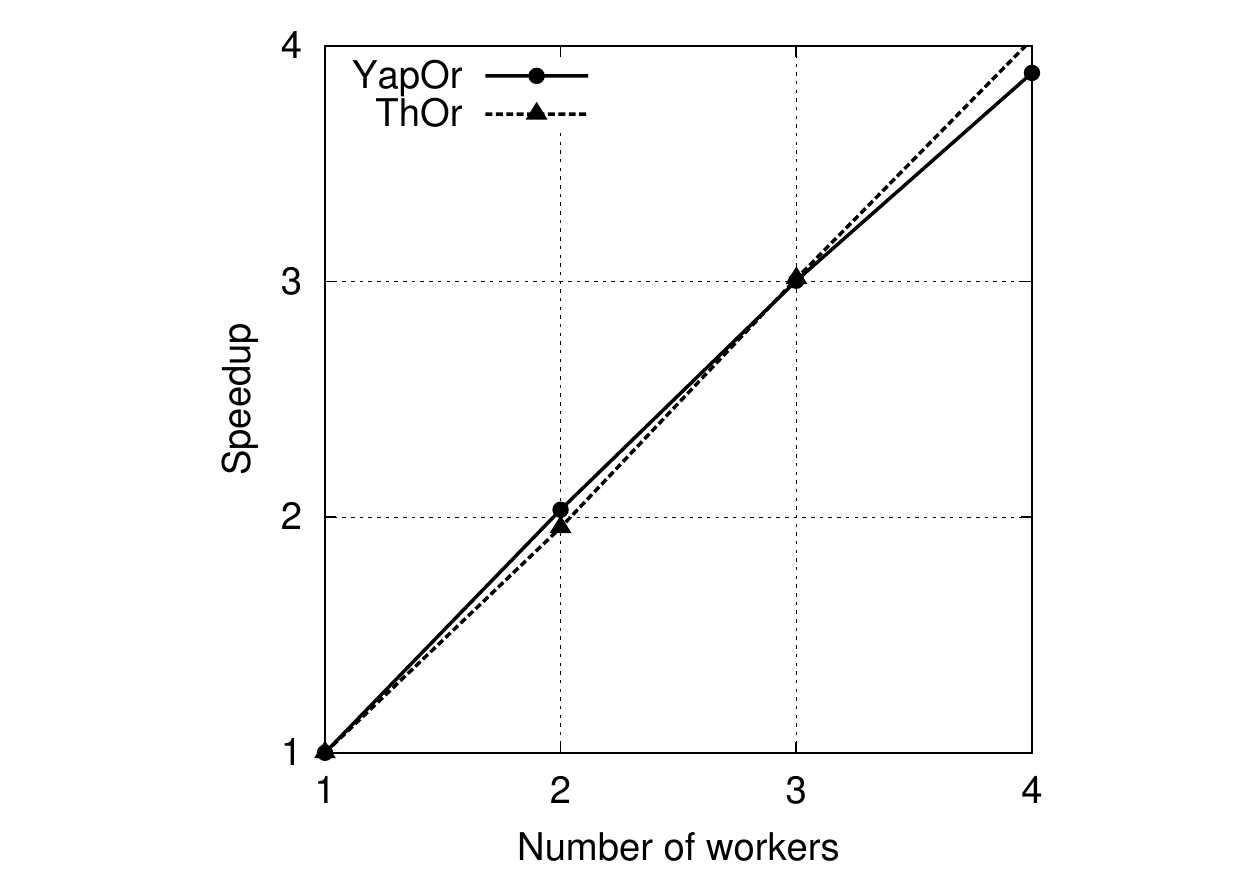}\hspace{-1cm}}
\subfigure[\tt map]{\label{fig:map_4}
          \hspace{-1cm}\includegraphics[width=6cm]{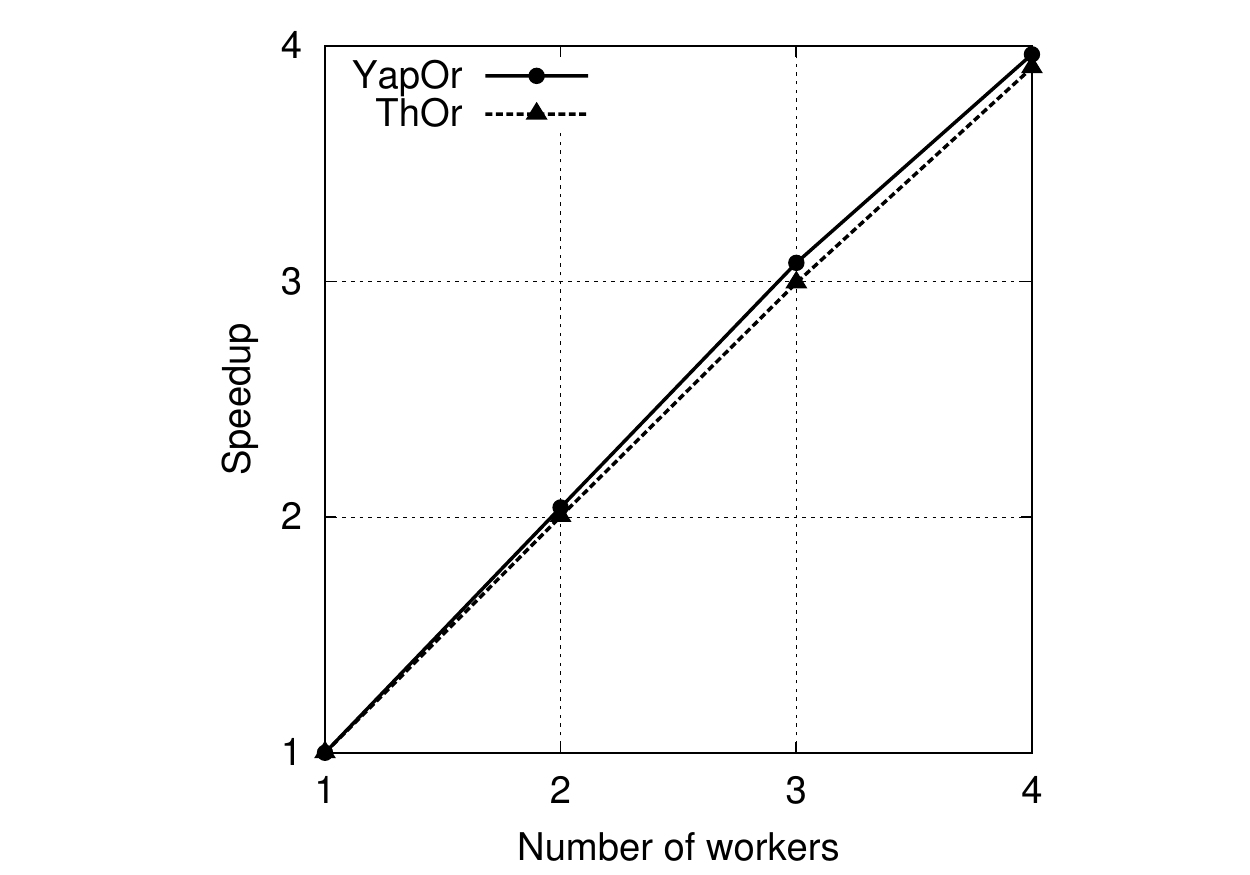}\hspace{-1cm}}
\subfigure[\tt mapbigger]{\label{fig:mapbigger_4}
          \hspace{-1cm}\includegraphics[width=6cm]{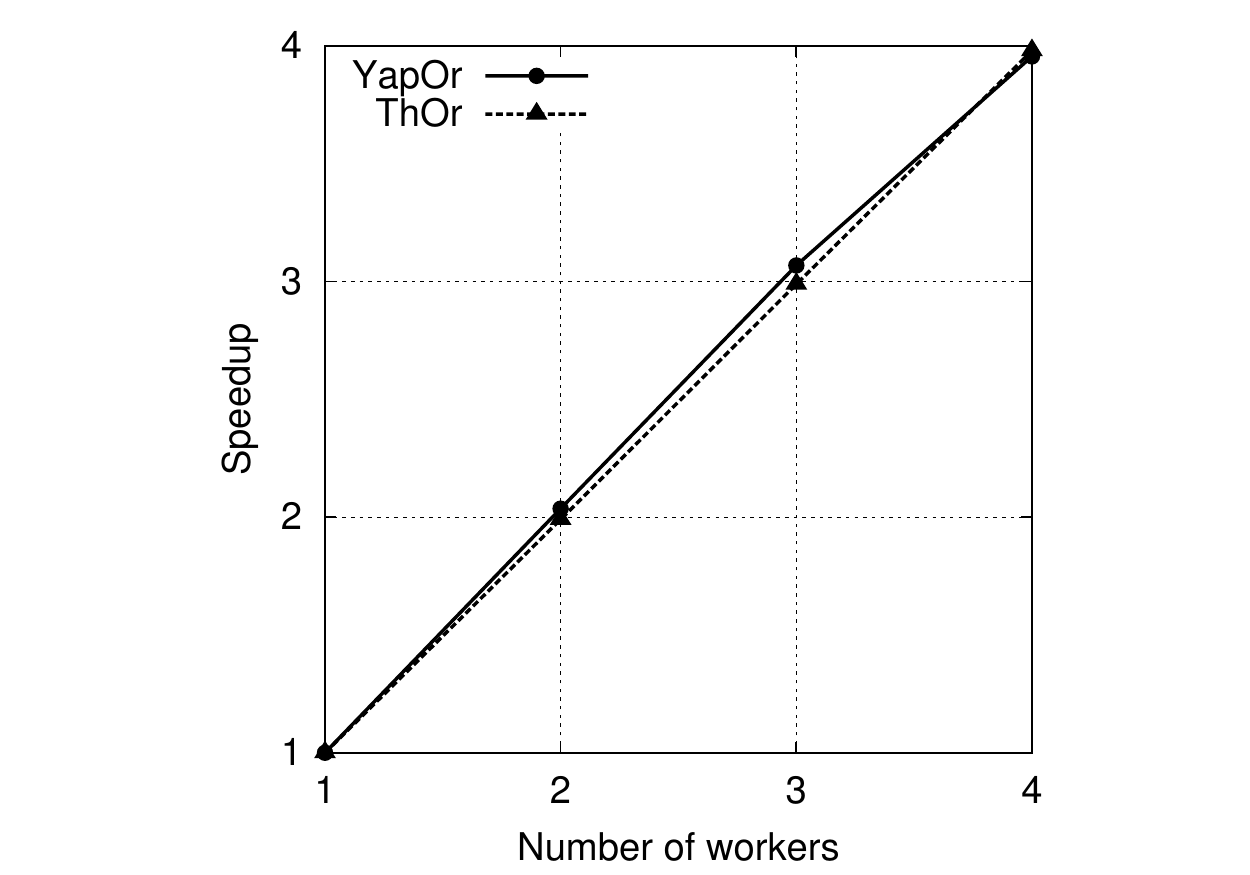}\hspace{-1cm}}
\subfigure[\tt puzzle]{\label{fig:puzzle_4}
          \hspace{-1cm}\includegraphics[width=6cm]{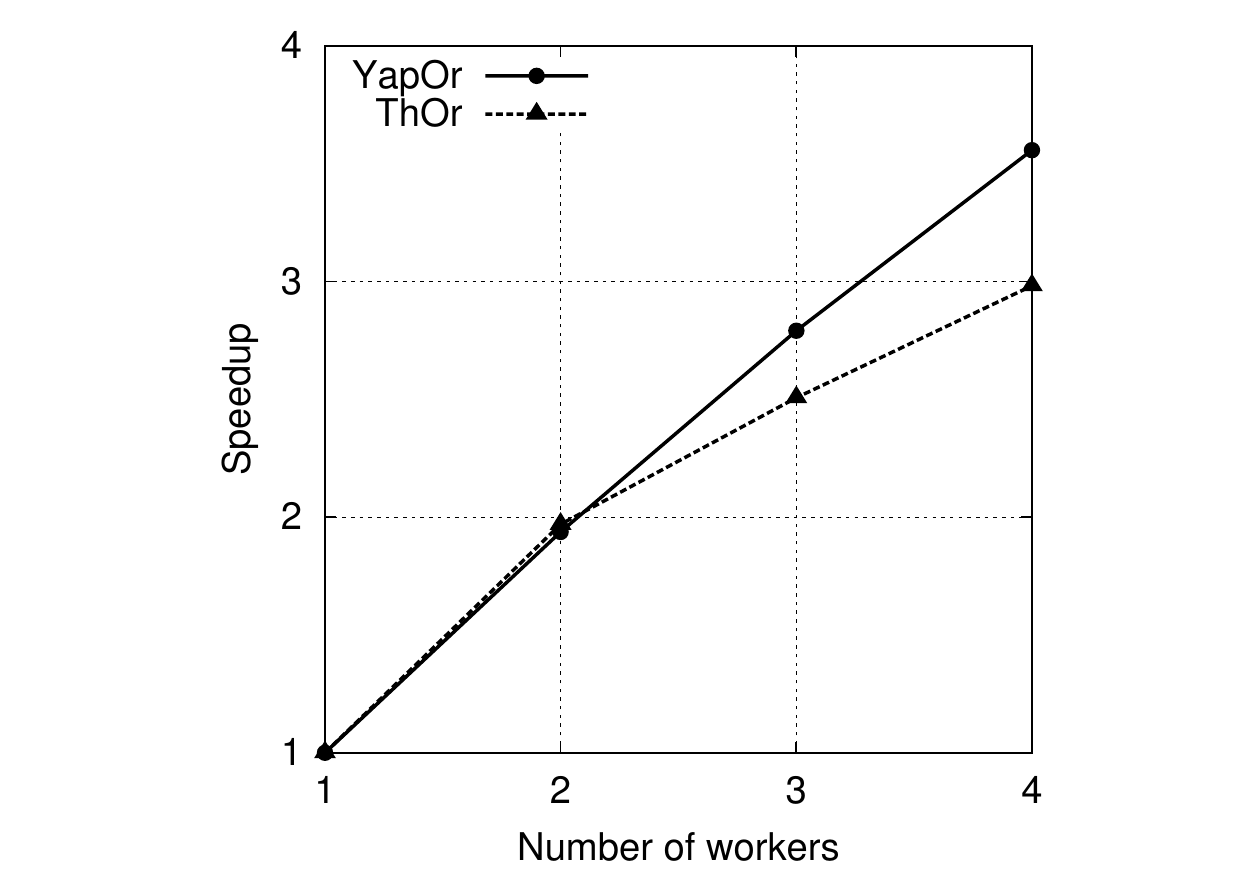}\hspace{-1cm}}
\subfigure[\tt puzzle4x4]{\label{fig:puzzle4x4_4}
          \hspace{-1cm}\includegraphics[width=6cm]{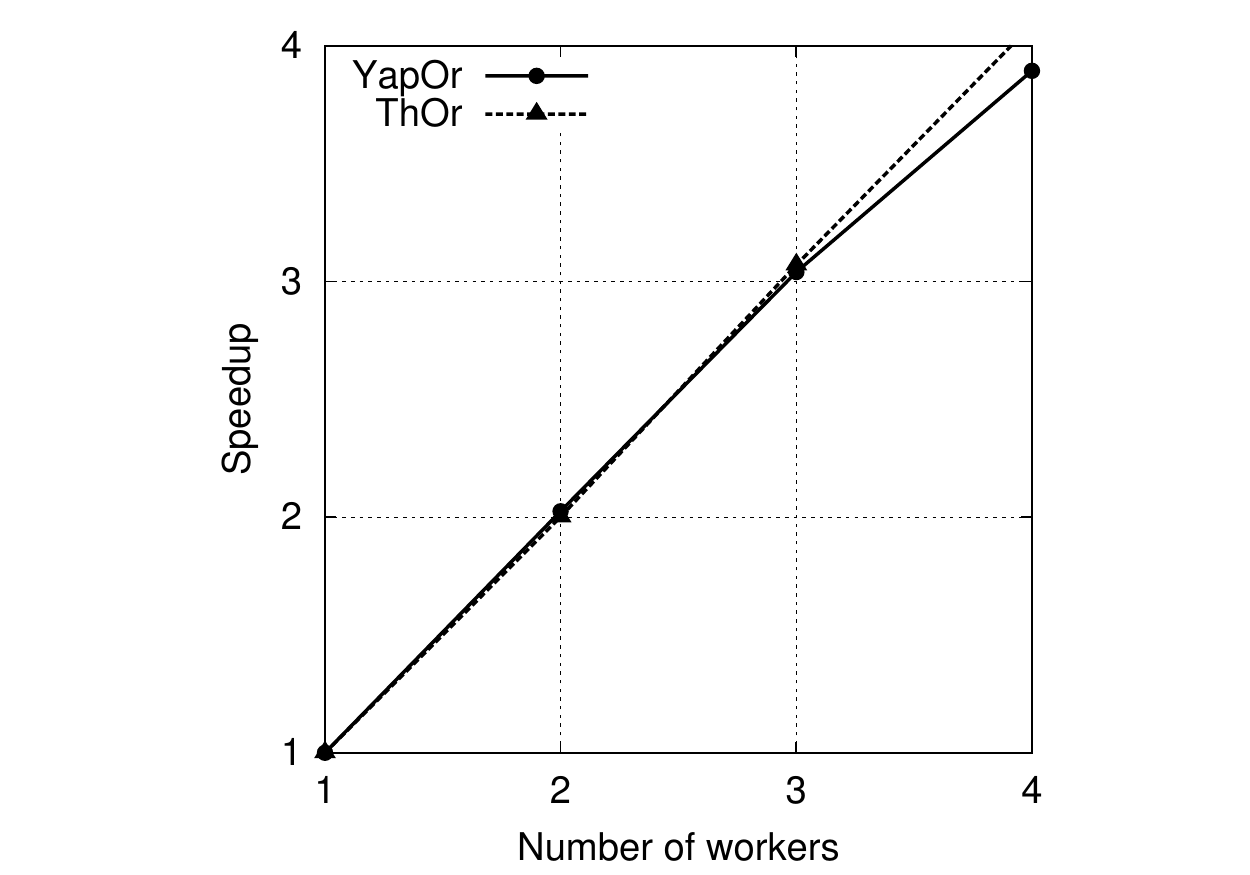}\hspace{-1cm}}
\subfigure[\tt queens]{\label{fig:queens_4}
          \hspace{-1cm}\includegraphics[width=6cm]{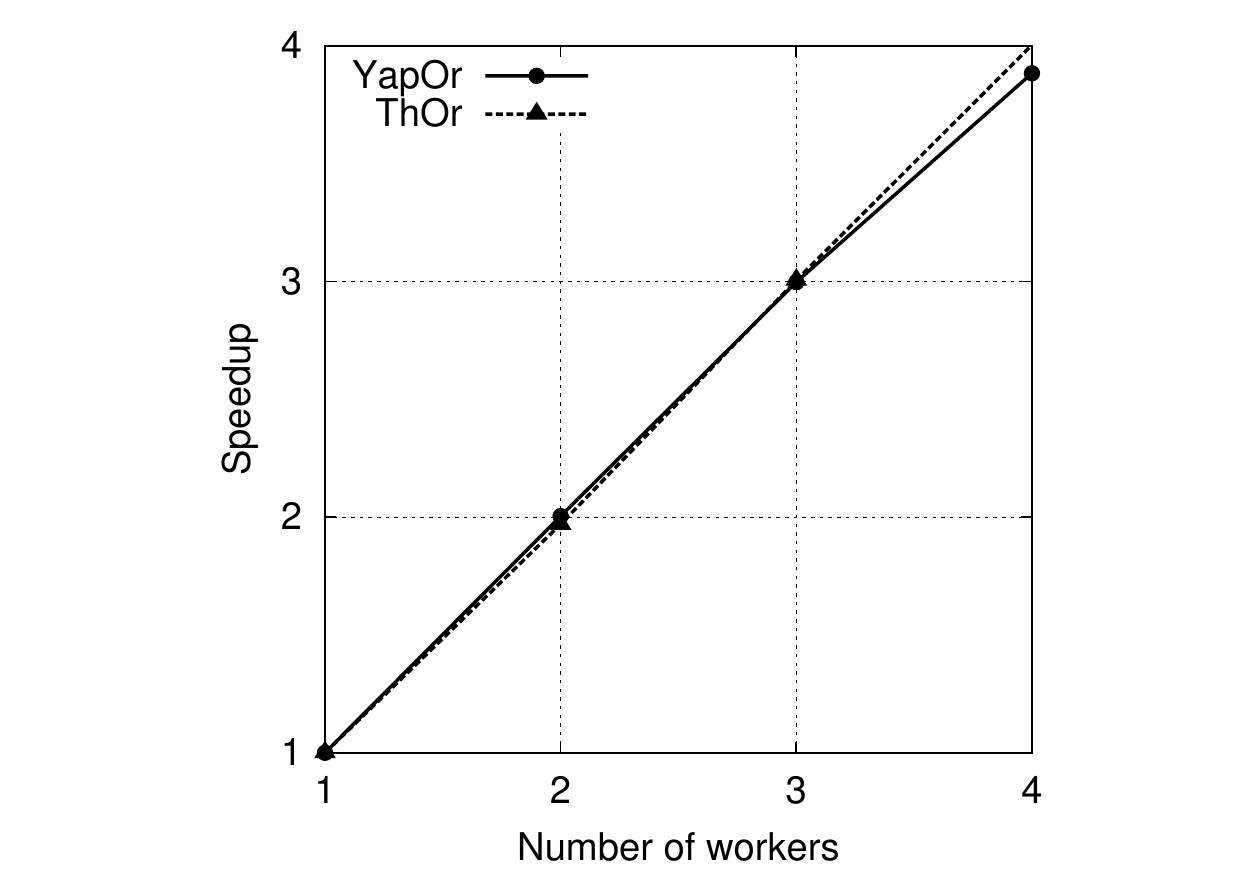}\hspace{-1cm}}
\caption{Speedups for benchmark applications on the Linux quad-core desktop
  machine}
\label{fig:speedups_desktop}
\end{figure}

\begin{figure}
\subfigure[\tt cubes]{\label{fig:cubes_24}
          \hspace{-1cm}\includegraphics[width=6cm]{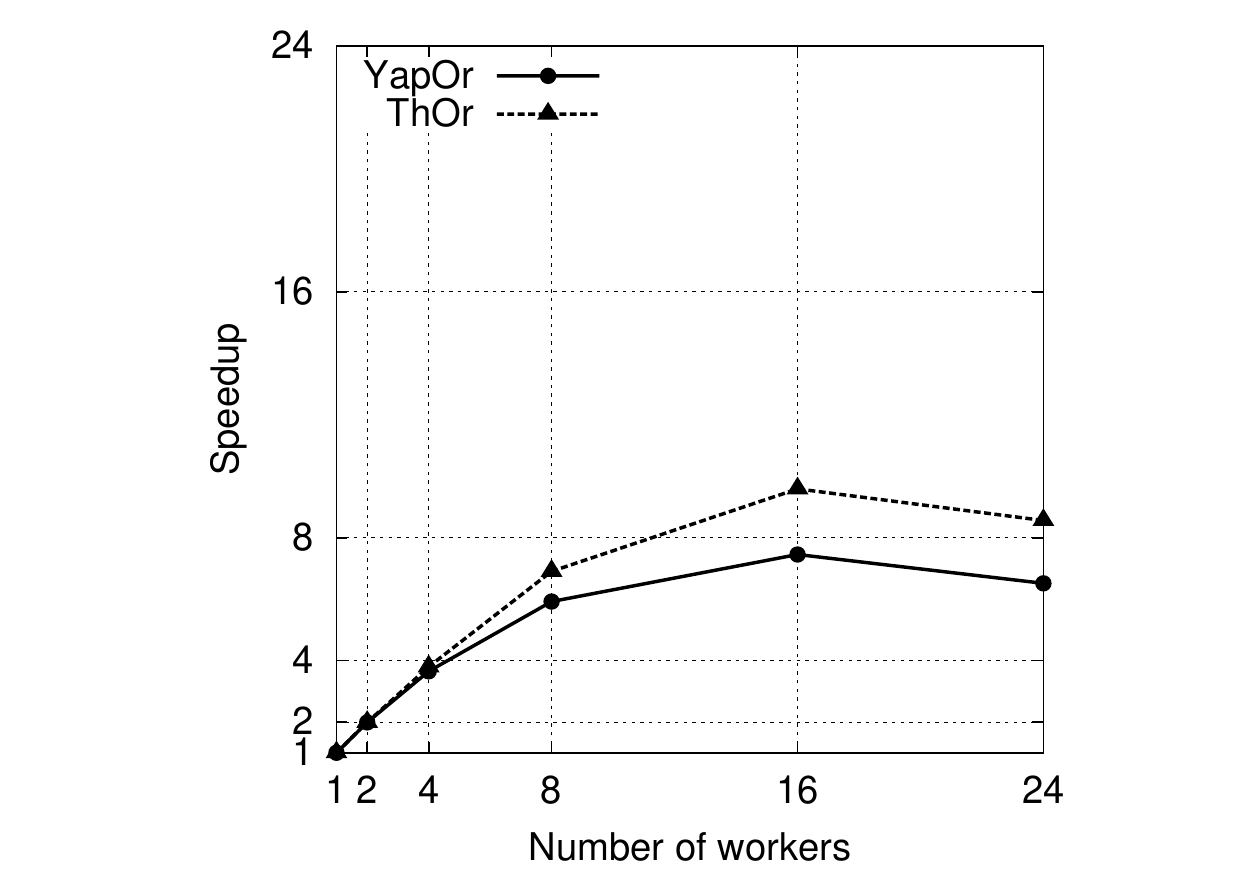}\hspace{-1cm}}
\subfigure[\tt fp]{\label{fig:fp_24}
          \hspace{-1cm}\includegraphics[width=6cm]{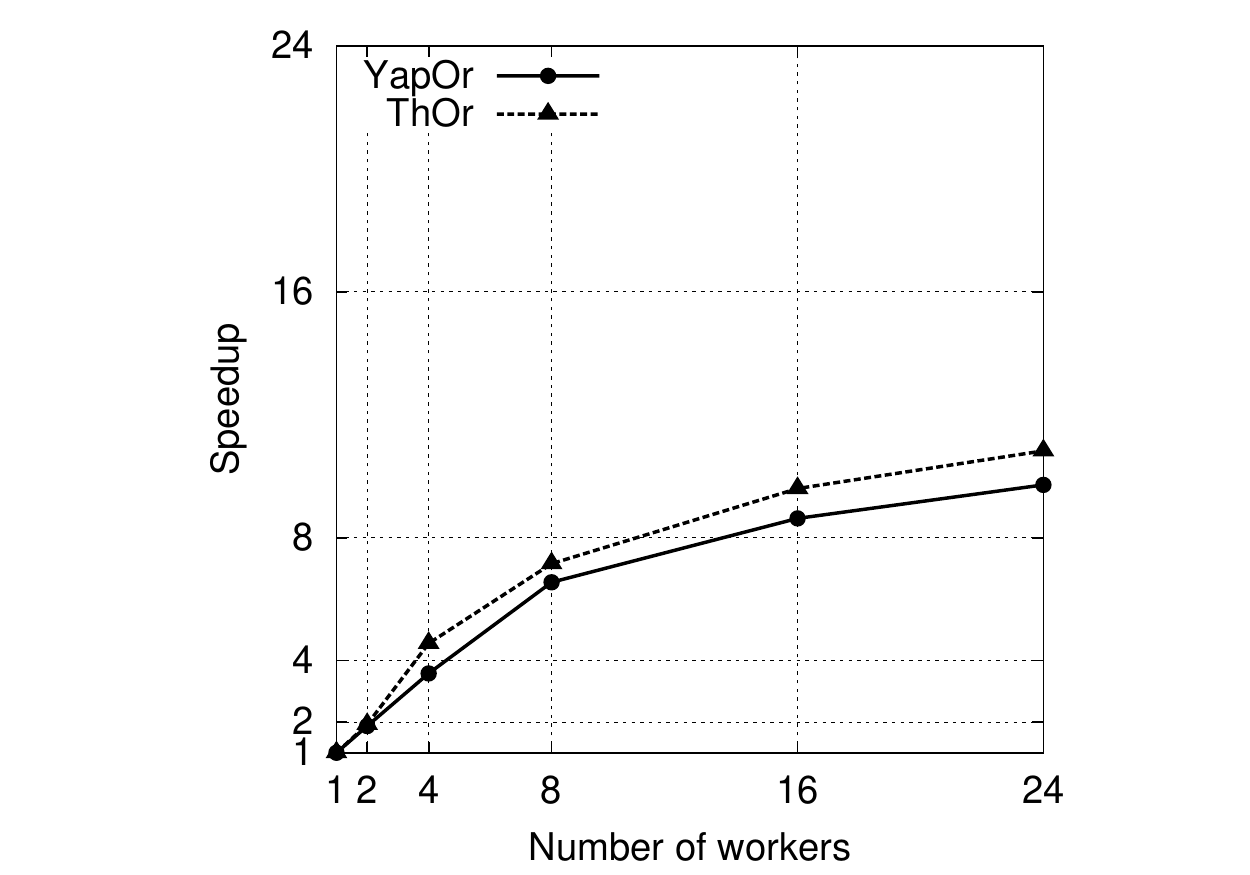}\hspace{-1cm}}
\subfigure[\tt ham]{\label{fig:ham_24}
          \hspace{-1cm}\includegraphics[width=6cm]{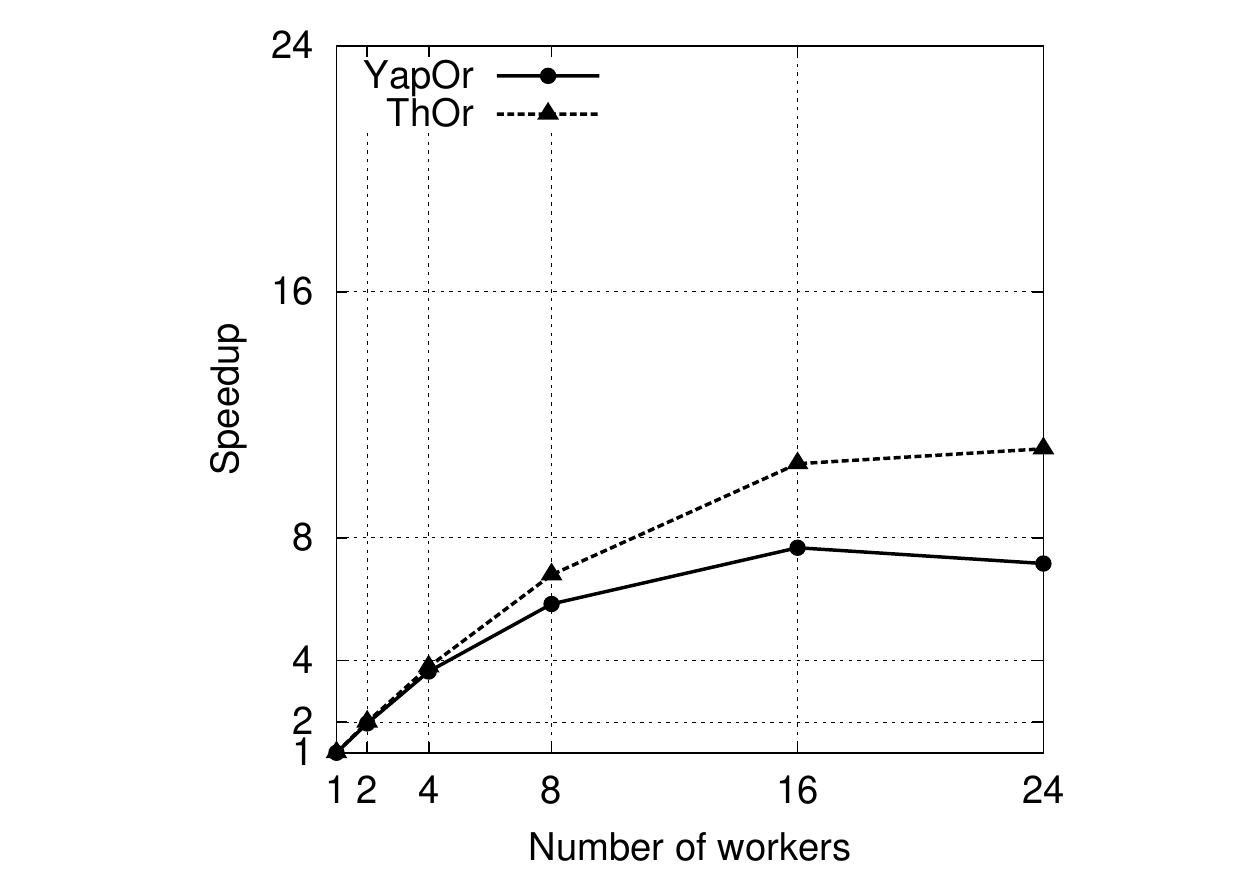}\hspace{-1cm}}
\subfigure[\tt magic]{\label{fig:magic_24}
          \hspace{-1cm}\includegraphics[width=6cm]{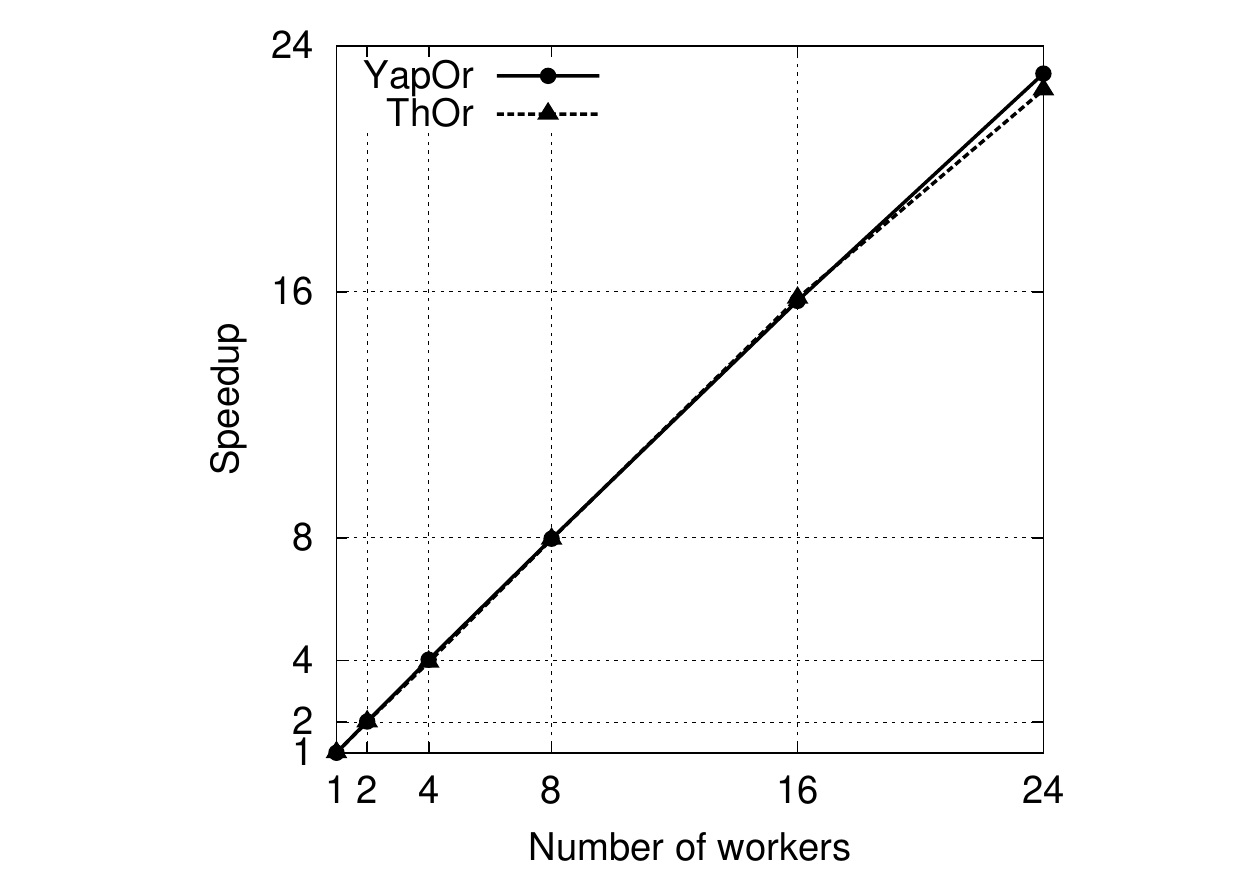}\hspace{-1cm}}
\subfigure[\tt map]{\label{fig:map_24}
          \hspace{-1cm}\includegraphics[width=6cm]{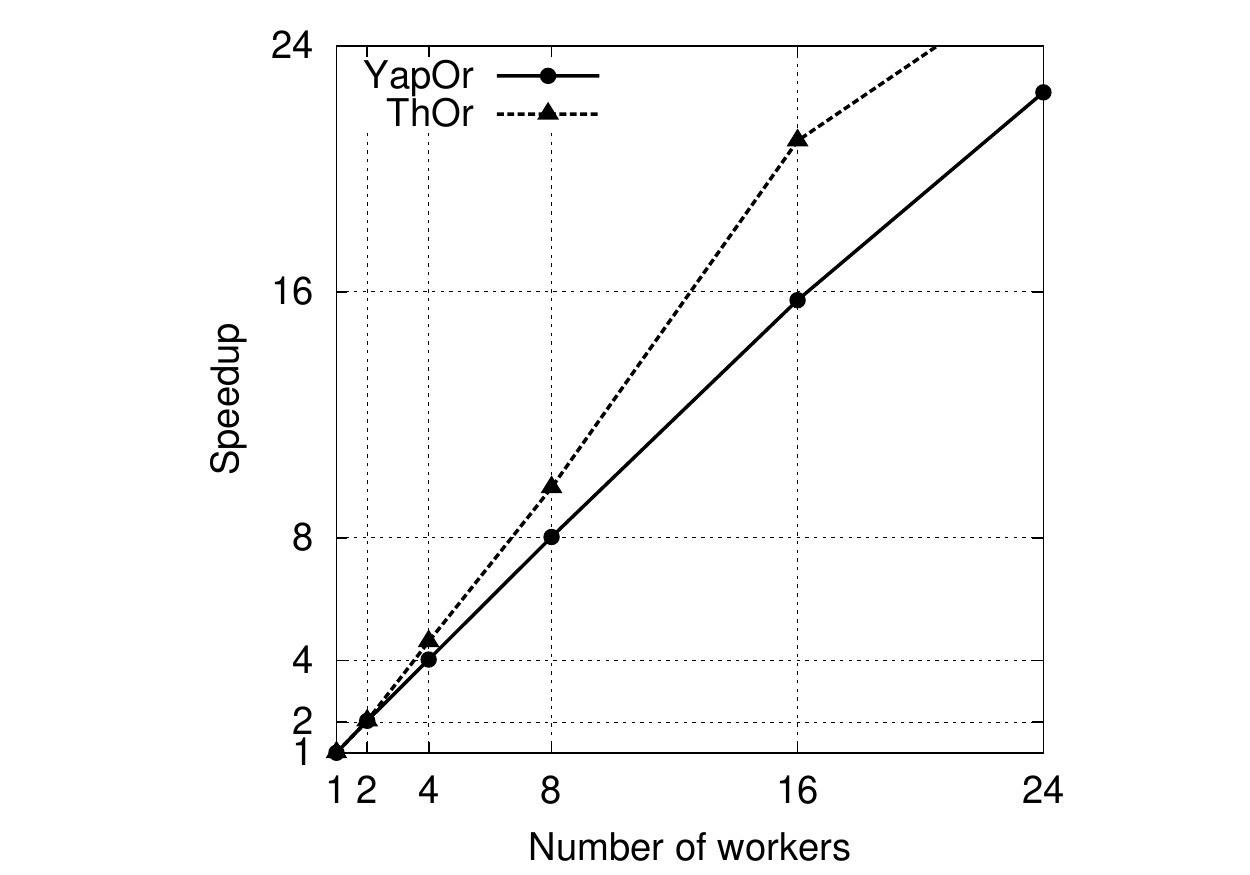}\hspace{-1cm}}
\subfigure[\tt mapbigger]{\label{fig:mapbigger_24}
          \hspace{-1cm}\includegraphics[width=6cm]{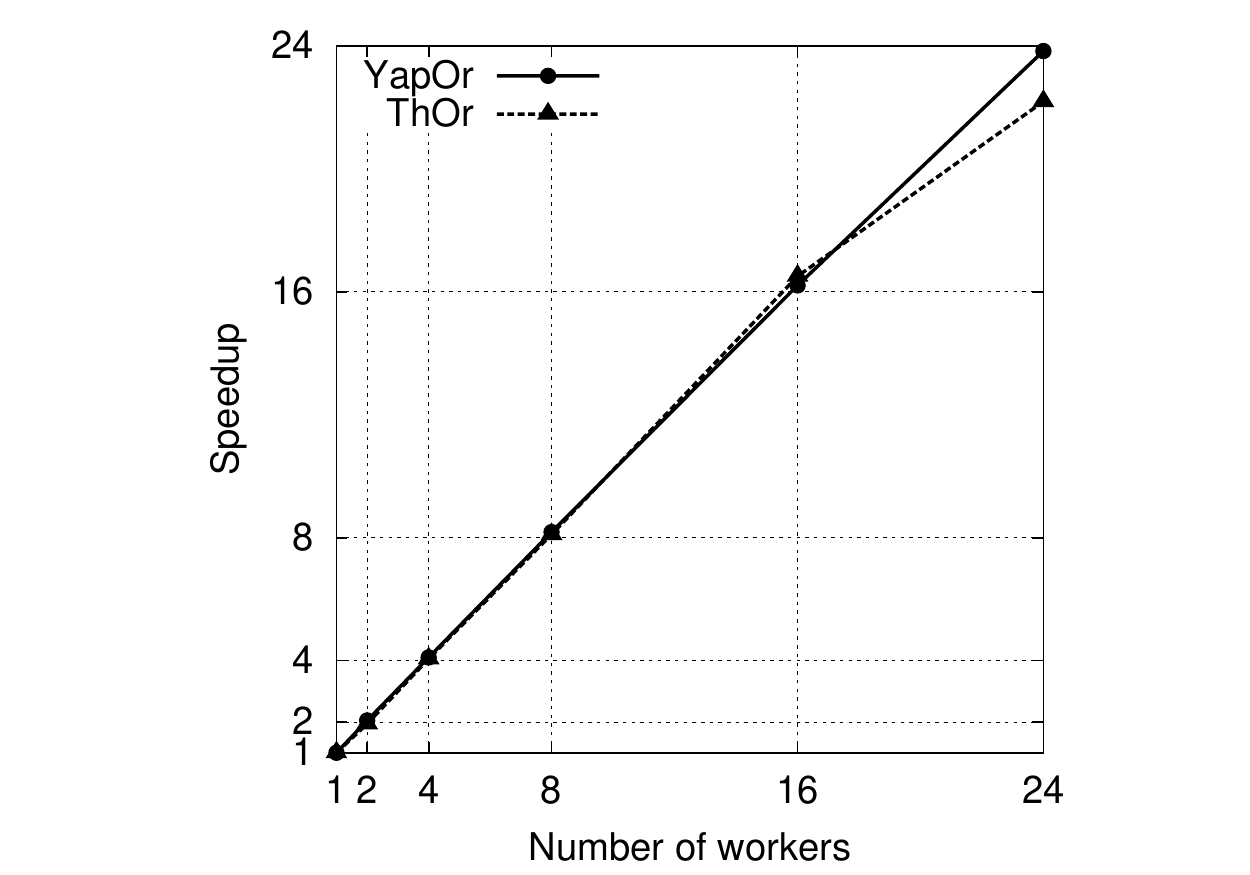}\hspace{-1cm}}
\subfigure[\tt puzzle]{\label{fig:puzzle_24}
          \hspace{-1cm}\includegraphics[width=6cm]{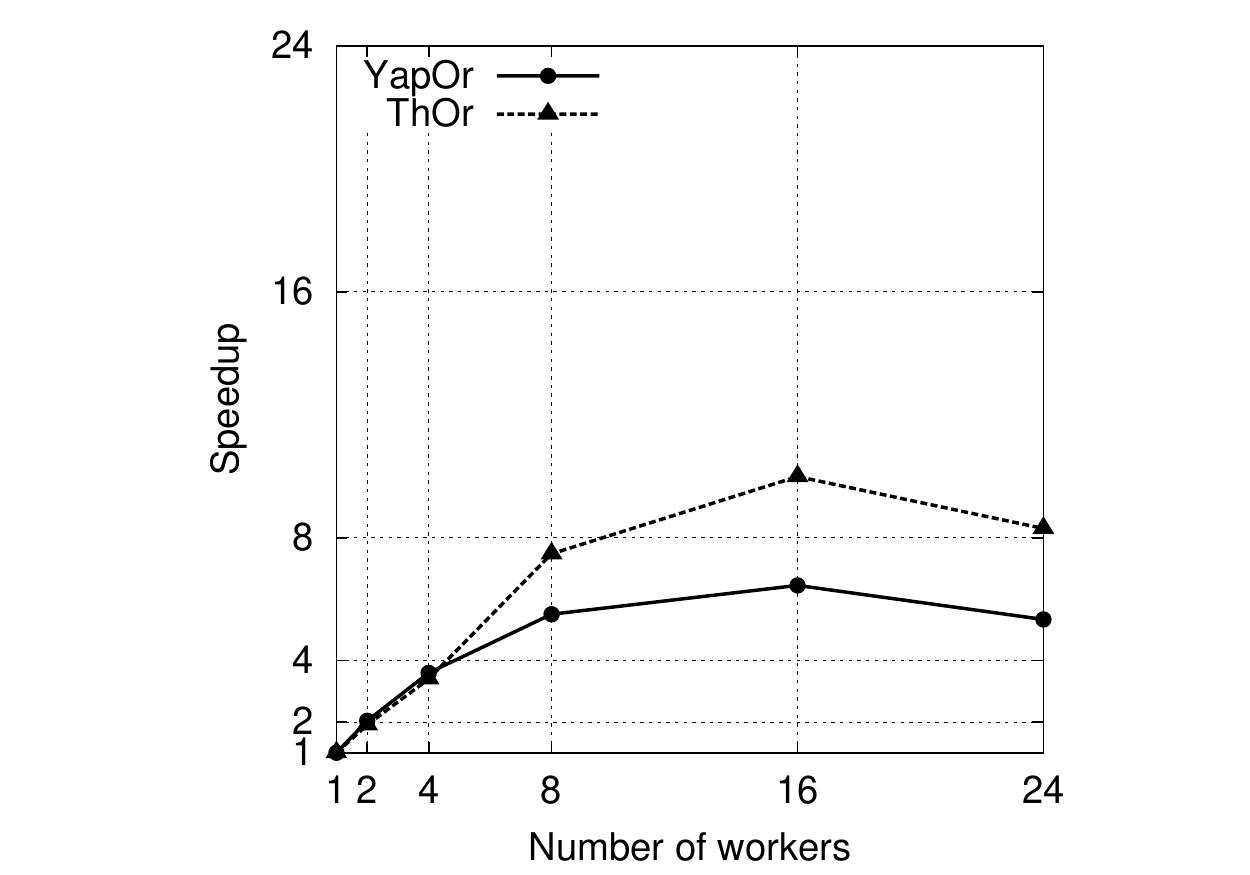}\hspace{-1cm}}
\subfigure[\tt puzzle4x4]{\label{fig:puzzle4x4_24}
          \hspace{-1cm}\includegraphics[width=6cm]{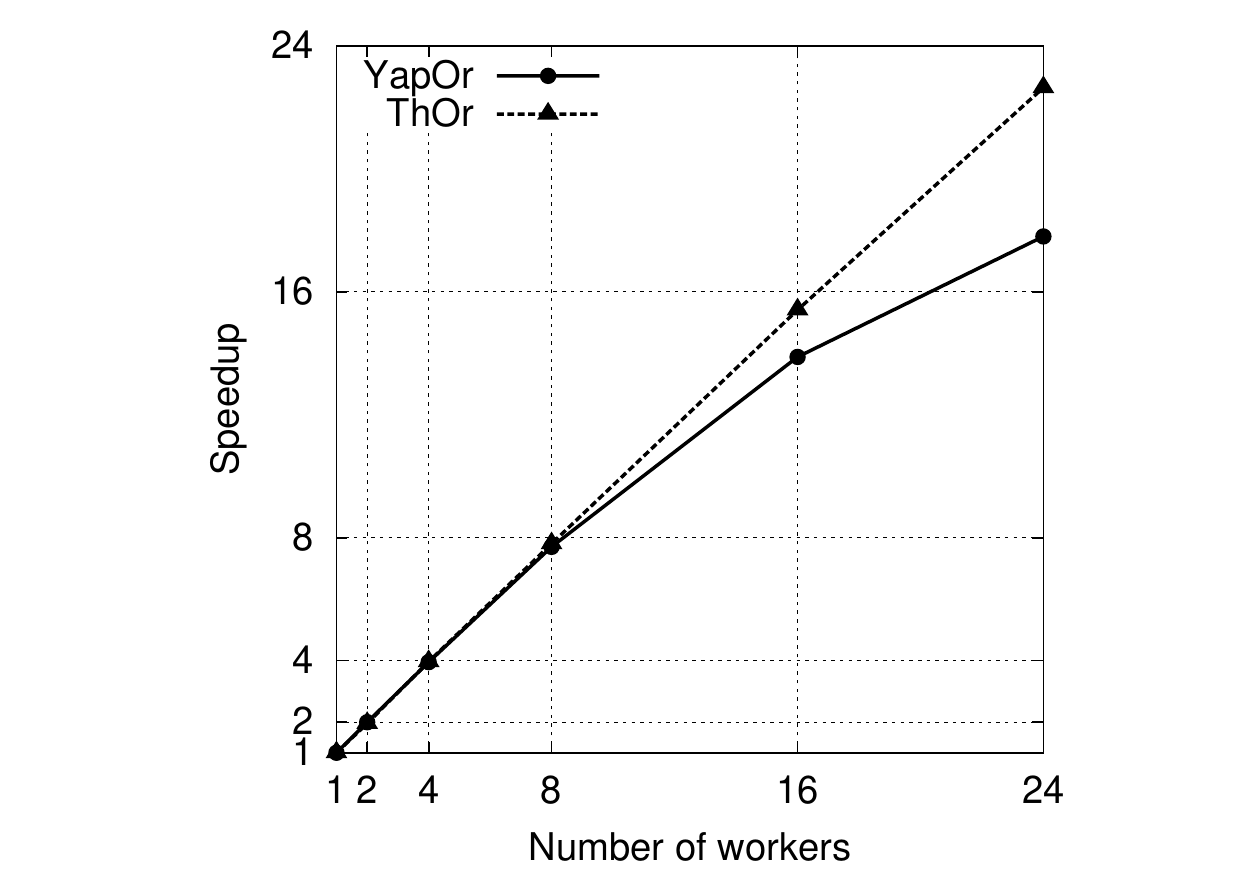}\hspace{-1cm}}
\subfigure[\tt queens]{\label{fig:queens_24}
          \hspace{-1cm}\includegraphics[width=6cm]{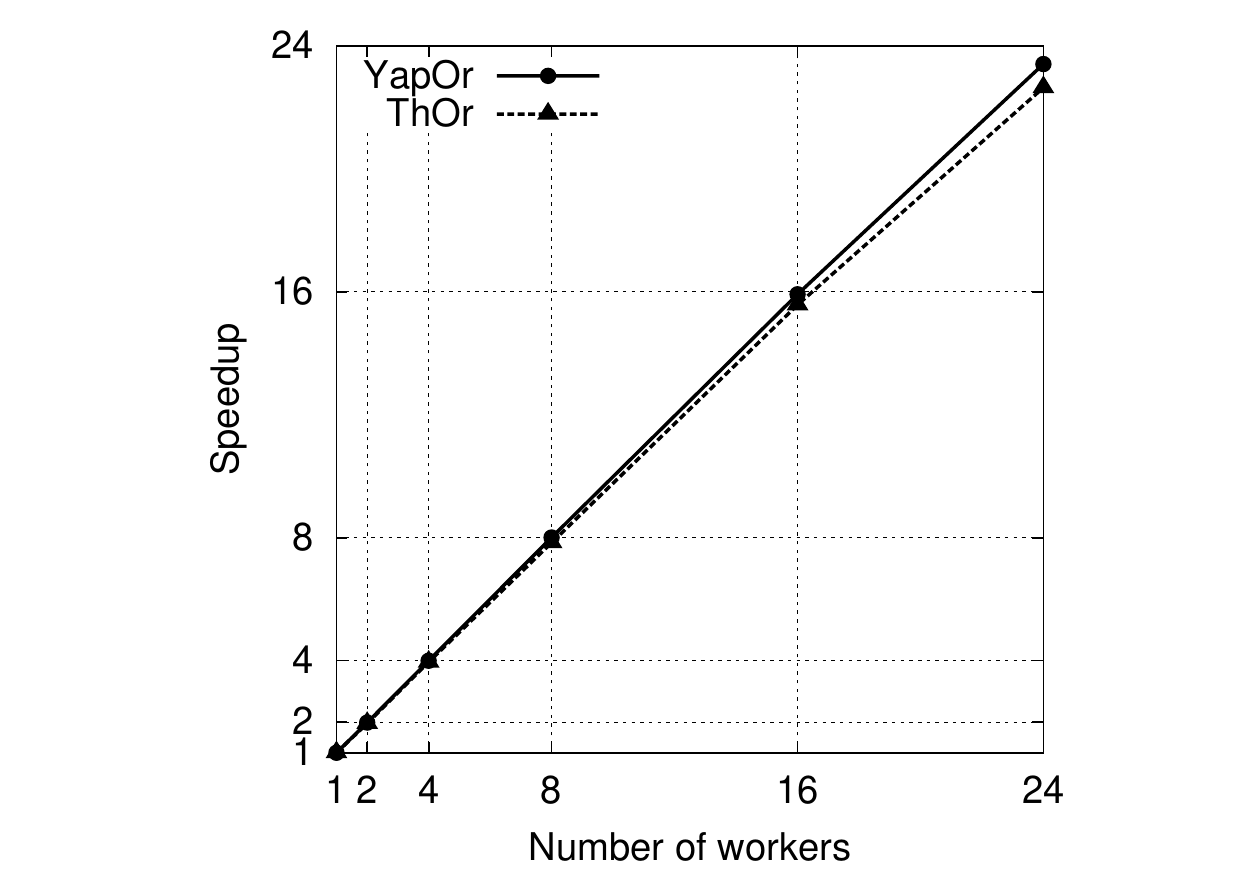}\hspace{-1cm}}
\caption{Speedups for benchmark applications on the Linux 4 six-core
  server machine}
\label{fig:speedups_server}
\end{figure}

Figures~\ref{fig:speedups_desktop} and~\ref{fig:speedups_server} compare the speedups
of YapOr and ThOr for each benchmark in the quad-core and in the 4
six-core server. We achieve quasi-ideal speedups for almost all
benchmarks in the quad-core machine. In the 4 six-core server, ThOr
and YapOr manage to achieve quasi-linear speedups in all applications
that have enough parallelism for 24 processors. Amongst the real-world
applications, \texttt{fp} achieves very good speedups even for 24
processors. It is interesting to notice that ThOr is capable of
extracting parallelism and achieving good speedups even from the
smallest benchmarks.

Obviously, for larger numbers of cores, one may eventually observe a
drop in performance as is the case of \texttt{fp}, in the quad-core
machine, with 4 processors (Figure~\ref{fig:fp_4}).

The threaded version may produce results such as the ones observed for
the application {\tt map}, in Figure~\ref{fig:map_24}, where it
achieves super-linear speedups. The application has no pruning, and
the speedups tend to improve as we add more cores. This suggests the
speedup may be memory related, and caused by cache effects.

It is also interesting to show performance results on a dual-core
laptop. In this case, we use an Apple MacBook Pro 2.5 GHz Intel Core 2
Duo with 4 GBytes of RAM. The machine can run two operating systems:
Mac OS X version 10.6.2 and Windows Vista. We experiment with both
operating systems on a relatively idle machine (although not in single
user) and report average run-time for one and two workers (please see
Table~\ref{tab:non_unix}). Notice that YapOr requires Unix, so it
could not run in either of these two configurations.

\begin{table}[hbtp]
\caption{Average execution times, in seconds, for one and two cores of
  ThOr on an Apple laptop machine running OS X and Windows Vista}
\begin{tabular}{l|rr|rr} \hline
\multirow{2}{*}{\bf Benchmark} &
\multicolumn{2}{c|}{\bf OS X} & 
\multicolumn{2}{c}{\bf Windows Vista} \\ 
& \multicolumn{1}{c}{\bf 1} & \multicolumn{1}{c|}{\bf 2}
& \multicolumn{1}{c}{\bf 1} & \multicolumn{1}{c}{\bf 2} \\ 
\hline\hline 
\texttt{cubes}     &  0.103 &  0.056 (1.84) &  0.156 &  0.078 (2.00) \\
\texttt{fp}        &  1.930 &  1.040 (1.86) &  3.500 &  1.790 (1.96) \\
\texttt{ham}       &  0.220 &  0.110 (2.00) &  0.530 &  0.280 (1.89) \\
\texttt{magic}     & 27.500 & 15.000 (1.83) & 30.700 & 15.600 (1.97) \\
\texttt{map}       & 15.800 &  9.000 (1.76) & 33.500 & 17.700 (1.89) \\
\texttt{mapbigger} & 43.500 & 23.500 (1.85) & 92.900 & 49.000 (1.90) \\
\texttt{puzzle}    &  0.100 &  0.055 (1.82) &  0.220 &  0.110 (2.00) \\
\texttt{puzzle4x4} &  6.900 &  3.700 (1.86) &  8.980 &  4.600 (1.95) \\ 
\texttt{queens}    & 24.100 & 13.200 (1.83) & 29.700 & 15.000 (1.98) \\
\hline
\emph{Average}     &       &         (1.85) &        &        (1.94) \\
\hline
\end{tabular}
\label{tab:non_unix}
\end{table}

Our results clearly show significant speedups in exploiting
or-parallelism on modern multi-core architectures. In fact, we obtain
linear speedups for some applications up to very larger number of
cores. It is impressive that these speedups were based on what is a
twenty years old design. The key ideas, copying and bottom-most
sharing, seem to perform as well today as they did when they were
proposed.

A second observation is that there is a wide variation on the
baseline, depending on the operating system and execution mode. A
deeper analysis shows that this variation largely depends on execution
mode, compiler used, and support for threads. Support for threads in
YAP is quite expensive, because YAP was designed around global
variables and multiple threads require the use of private variables
for each thread. On the other hand, we believe that ThOr overheads can
be easily reduced, and ThOr guarantees speedup on a wide range of
machines and operating systems.

We also observe that both ThOr and YapOr yield very close speedups for
almost all applications. This is partly because some of these
applications are relatively small, and incremental copying makes the
cost of copying and adjusting addresses a very minor component of the
cost equation. 
Indeed, in examples such as \texttt{queens}, the overheads
are not very noticeable because the parallel work is coarse-grained thus
minimizing the number of copies. For a run of 13 queens we measured
from 4 to 7 work sharing operations between two workers during the
run-time of approximately 20 seconds. On the other hand,  ThOr seems to
be doing better than YapOr even in fine grained benchmarks. This
deserves further investigation, but is probably
because its baseline performance is worse.

One important advantage of ThOr is that it relies only on thread
support: nowadays, POSIX thread support can be found even in Windows
systems, although the significant loss of performance we found
suggests POSIX thread implementations may not always be up to the
best.  The fact that ThOr can run in a wide variety of platforms, and
achieves good speedups, is a major advantage of this proposal over
previous systems.


\section{Conclusions and Further Work}

We presented ThOr, a novel system for the exploitation of
or-parallelism in logic programming systems. Our approach is based on
threads, but differs from other
proposals~\cite{Casas-08,Moura-08,Moura-09} in that, instead of
building a system from scratch by using a high-level approach, we
reuse as much possible the excellent previous work in parallel logic
programming. Our results indicate that our approach has three main
advantages: it allows excellent scalability, and the trend is to
expect more cores per computer; it allows good performance on
fine-grained tasks, with such speedups being quite useful for users
running laptops or personal desktops; it provides an excellent
parallel engine with dynamic scheduling for irregular applications.

We can see a large number of future paths for ThOr. One obvious path
is to make the system more robust and to experiment with different
scheduling strategies, such as stack-splitting~\cite{Pontelli-06}. A
second direction is to experiment with more applications, namely with
tabled programs and with constraint programs. We believe our task will
be simplified by the fact that YAP already supports a number of
constraint solvers. Last, but not least, we believe our approach gives
logic programming a strategy to combine both implicit and explicit
parallelism. Our goal is to have specialized threads that can do
input/output or that can just take care of sequential execution and
offload parallelizable tasks to or-parallel servers. We believe that
such a flexible approach will be a key for parallel logic programming
to fulfill its great promise.


\section*{Acknowledgments}

This work has been partially supported by the FCT research projects
STAMPA (PTDC/EIA/67738/2006) and HORUS (PTDC/ EIA-EIA/100897/2008). We
would like to thank all the anonymous referees that helped improving
the quality of this paper with their insightful comments and
suggestions.


\bibliographystyle{acmtrans}
\bibliography{references}

\end{document}